\documentclass[11pt]{article}

\usepackage[preprint]{acl}

\usepackage{times}
\usepackage{latexsym}

\usepackage[T1]{fontenc}
\usepackage[utf8]{inputenc}

\usepackage{microtype}

\usepackage{inconsolata}

\usepackage{graphicx}
\usepackage{amsmath}
\usepackage{amssymb}
\usepackage{booktabs}
\usepackage{multirow}
\usepackage{colortbl}
\usepackage{array}
\usepackage{algorithm}
\usepackage{algpseudocode}
\usepackage{tcolorbox}
\usepackage{caption}
\usepackage{xcolor}

\tcbuselibrary{skins}

\newcommand{\sysname}{EvoMemState}
\definecolor{rowblue}{HTML}{EAF3FF}
\newcolumntype{L}[1]{>{\raggedright\arraybackslash}p{#1}}
\newcolumntype{C}[1]{>{\centering\arraybackslash}m{#1}}

\title{EvoMemState: Evolving Memory State for Personal Agents}

\author{
  \textbf{Zhishang Xiang}$^{1}$\thanks{Equal contribution.},
  \textbf{Yunbo Tang}$^{1}$\footnotemark[1],
  \textbf{Zhongqiao Liu}$^{2}$,
  \textbf{Junkai Chen}$^{1}$,
  \textbf{Zerui Chen}$^{1}$,
  \\
  \textbf{Qingguo Hu}$^{1}$,
  \textbf{Yuxuan Zhang}$^{3}$,
  \textbf{William Grant}$^{4}$,
  \textbf{Qinggang Zhang}$^{5}$\thanks{Corresponding authors.},
  \textbf{Jinsong Su}$^{1}$\footnotemark[2]
  \\[2pt]
  $^{1}$Xiamen University
  \qquad
  $^{2}$Nanyang Technological University
  \\
  $^{3}$University of British Columbia
  \qquad
  $^{4}$University of Waterloo
  \qquad
  $^{5}$Jilin University
  \\[2pt]
  \texttt{\{xiangzhishang,tangyunbo\}@stu.xmu.edu.cn}
  \\
  \texttt{qinggangzhang@jlu.edu.cn}
  \qquad
  \texttt{jssu@xmu.edu.cn}
}
\begin{document}
\maketitle
\begin{abstract}
Long-term memory enables large language model (LLM) agents to provide continuous personalized assistance by retaining user information across interactions. Existing systems support this process through fact extraction, organization, updating, and retrieval, but largely maintain memories as independent facts. This fact-centric paradigm may overlook how new facts update, constrain, or conflict with earlier ones, while leaving their applicability, role, and priority for the current request implicit. In this paper, we introduce \textbf{EvoMemState}, a state-centric framework that augments persistent facts with two complementary states: Relational States that capture how facts remain valid and relate to one another as memory evolves, and a Reasoning State that determines how these facts should be applied to the current request. Specifically, \textsc{EvoMemState} consists of three key components: (i) Ontology-Guided Memory State Induction extracts Persistent Facts and constructs Relational States that maintain their validity, scope, and evolving relations. (ii) RL-Based State Evolution internalizes memory-use experience into a policy through reinforcement learning to generate a request-specific Reasoning State. (iii) State-Aware Reasoning then combines the relevant facts with their Relational States and the generated Reasoning State to guide the personal agent. Extensive experiments on long-term dialogue and memory-use benchmarks demonstrate consistent improvements over existing memory systems in retrieval completeness and memory-grounded response accuracy. Our code and implementation details are available at \textcolor{blue}{\url{https://github.com/XMUDeepLIT/EvoMemState.git}}.
\end{abstract}

% This motivates a state-centric paradigm in which memory states capture the validity, scope, and relations of facts and specify how they should be used in the current interaction. Following this paradigm, we introduce \textsc{EvoMemState} for maintaining and using long-term personal memory. 

\section{Introduction}

Large language model (LLM) agents are increasingly used as personal agents that assist users over extended periods, requiring continuity across interactions and adaptation to evolving needs~\citep{westhausser2025enabling,zhang2025survey,hu2025memory,feng2026graph}. Long-term memory supports this continuity by preserving user preferences, experiences, circumstances, and constraints across sessions~\citep{li2026survey,du2025rethinking,ning2026memadapter,yang2026graph}. This accumulated context provides a foundation for consistent and personalized assistance over time~\citep{wang2026assistant,zhang2026memsim}.

Recent work has explored how to construct and maintain long-term personal memory from interaction histories. MemoryBank~\citep{zhong2023memorybank} summarizes user experiences for subsequent retrieval, while MemGPT~\citep{packer2023memgpt} manages memory across storage tiers. LightMem~\citep{wu2025lightmem} improves efficiency through compression and selective retrieval. Other systems emphasize memory organization and maintenance: Mem0~\citep{chhikara2025mem0} extracts and updates atomic facts, A-MEM~\citep{xu2025amem} dynamically associates related memories, and Supermemory~\citep{supermemory2026} and SimpleMem~\citep{liu2026simplemem} support continual updates and memory lifecycle management. Together, these approaches enable personal agents to retain and access information across interactions through memory extraction, organization, updating, and retrieval.

% 尽管已有研究取得了这些进展，存储的个人信息与个人智能体的实际行为之间仍然存在两个缺口。第一，\textbf{记忆状态演化缺口（Memory-State Evolution Gap）}：分散的事实记录和局部更新无法自动形成对用户的完整状态，也难以根据记忆在后续任务中的使用效果持续改进该状态。第二，\textbf{状态驱动推理缺口（State-Grounded Reasoning Gap）}：即使系统已经维护了用户状态，也仍需在面对新请求时选择相关的事实和状态，并确保推理过程真正建立在这些信息之上。因此，个人智能体可能保存了正确的用户事实，却没有形成准确的当前状态；也可能拥有相关状态，却没有在当前任务中正确使用它。
%
Despite this progress, existing memory systems largely follow a \textbf{fact-centric paradigm}, focusing on the storage, updating, and retrieval of factual records. However, the validity and scope of a stored fact may change as related memories accumulate, while its appropriate use depends on the current request. Associating records or revising their contents alone may not adequately capture these dependencies. When memories are maintained or retrieved separately, conflicts between them may be overlooked, leaving individually accurate records without the context needed for correct interpretation. Even when all relevant memories are retrieved, the personal agent must still determine their applicability and priority for the current request and may rely on outdated information or mishandle conflicts. These limitations motivate a \textbf{state-centric paradigm} that augments stored facts with evolving memory states. A memory state describes the status of stored facts and their roles in an interaction, which cannot be determined from factual content alone and must be considered in relation to other memories and the current request. Such states make the validity, scope, and relations of facts explicit and guide how retrieved memories should be used for the current request.

% For example, an earlier memory may indicate that the user enjoys mountain hiking, while later memories record a knee injury and a new preference for low-impact travel. Retrieving only the earlier preference leaves the agent with an incomplete understanding of the user; retrieving all these memories but still recommending a strenuous hike means that they have been used incorrectly. Together, these gaps allow incomplete or misapplied memories to shape the personal agent’s response, undermining its reliability as the user’s needs and circumstances evolve.

\begin{figure}[t]
    \centering
    \includegraphics[width=0.5\textwidth]{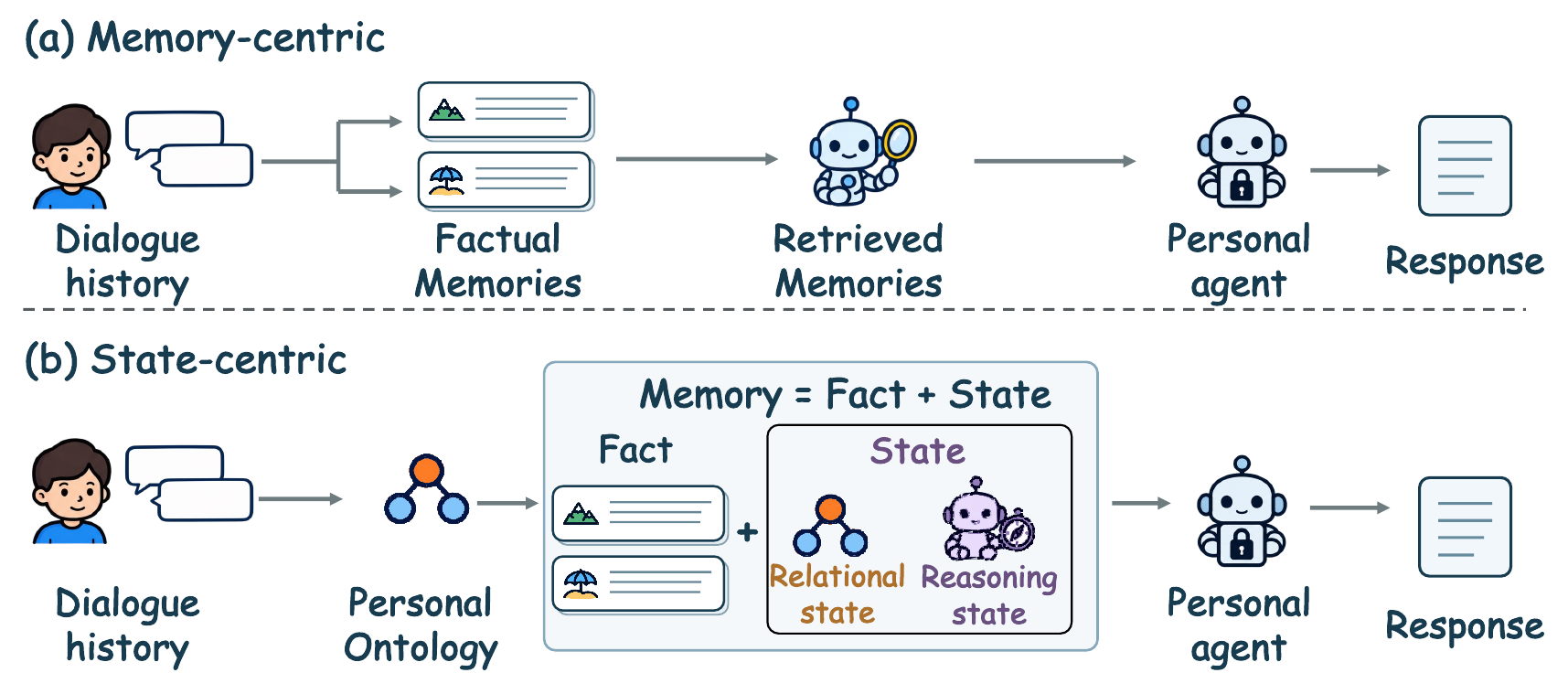}
    % \caption{Two gaps between memory and personal agent responses. The Memory-State Evolution Gap prevents records from forming an understanding of the user's current state, while the State-Grounded Reasoning Gap prevents that state from guiding the response.}
    \caption{Memory-centric and state-centric paradigms. Our retrieval preserves updates and conflicts, while generated reasoning states specify memory applicability, role, and priority for the current request.}
    \label{fig:intro}
    \vspace{-3mm}
\end{figure}

Augmenting personal memory with explicit states presents two challenges: (i) \textbf{It is challenging to construct and maintain coherent memory states.} Personal information accumulates across conversations as scattered and inconsistent observations. Simply collecting these observations leaves their relations implicit, while merging them into a summary may lose historical information and hide conflicts between records. The challenge is to organize related facts and maintain their validity, scope, and relations as new information arrives, while preserving the original records. (ii) \textbf{It is challenging to learn memory-use strategies from experience.} The same memory can serve different purposes across requests, so its validity and relations alone do not determine how it should affect the agent's behavior. This requires learning when memories should inform, constrain, or have no bearing on that behavior, using feedback from memory use to refine these judgments.

To address these challenges, we introduce \textsc{\sysname}, a state-centric framework for maintaining and using long-term personal memory. First, \textbf{Ontology-Guided Memory State Induction} extracts Persistent Facts from personal interactions and incrementally organizes them into a personal memory ontology. It preserves the original facts while maintaining their validity, scope, and evolving relations as Relational States. Second, \textbf{RL-Based State Evolution} internalizes memory-use experience into a policy through reinforcement learning. The learned policy generates a request-specific Reasoning State that specifies the applicability, role, and priority of relevant facts. Third, \textbf{State-Aware Reasoning} combines the relevant Relational States with the generated Reasoning State to guide the personal agent's response. Together, these components connect fragmented facts through Relational States and combine them with a request-specific Reasoning State, enabling personal agents to interpret and use memory appropriately.
% To address these challenges, we introduce \textsc{\sysname}, a state-centric framework that organizes long-term memory into state induction, state evolution, and state-aware reasoning. Rather than directly passing retrieved facts to a personal agent, \textsc{\sysname} processes memory through three complementary stages. First, \textbf{Ontology-Guided Memory State Induction} transforms accumulated personal information into Persistent Facts and constructs their Relational States to capture validity, scope, and evolving relations. Second, \textbf{RL-Based State Evolution} internalizes reusable memory-use experience through reinforcement learning and generates a request-specific Reasoning State that specifies the applicability, role, and priority of relevant facts. Third, \textbf{State-Aware Reasoning} combines the relevant Relational States with the generated Reasoning State to ground the personal agent in the current request. Together, these components transform isolated facts into Relational States and connect them to a request-specific Reasoning State, enabling personal agents to interpret and use memory appropriately.
Our contributions are summarized as follows:
\begin{itemize}
    \setlength{\itemsep}{2pt}
    \setlength{\parsep}{0pt}
    \setlength{\topsep}{3pt}

\item We identify the limitations of fact-centric personal memory: individually maintained facts may lack the relational context needed for interpretation, and complete retrieval does not ensure appropriate memory use.

\item We introduce \textsc{\sysname}, a state-centric long-term memory framework representing personal memory through Persistent Facts, Relational States, and Reasoning State. It realizes this formulation through Ontology-Guided Memory State Induction, RL-Based State Evolution, and State-Aware Reasoning.

\item We evaluate \textsc{\sysname} on long-term dialogue and memory-use benchmarks, demonstrating consistent improvements over existing memory systems in retrieval completeness and memory-grounded response accuracy.
\end{itemize}

\section{Preliminary Studies}

In this section, we conduct two preliminary studies on the Valid Memory Selection subset of MemSyco-Bench~\citep{xiang2026memsyco}, where correct answers require jointly considering memories linked by updates, conflicts, or constraints. Our findings reveal two limitations: related memories are not consistently retrieved together, leaving individual facts without the context needed for correct interpretation; and even when all relevant memories are retrieved, personal agents may misjudge their applicability, roles, and priorities for the current request. A detailed case analysis is provided in Appendix~\ref{app:preliminary-case}.

\begin{figure}[t]
    \centering
    \includegraphics[width=.95\columnwidth]{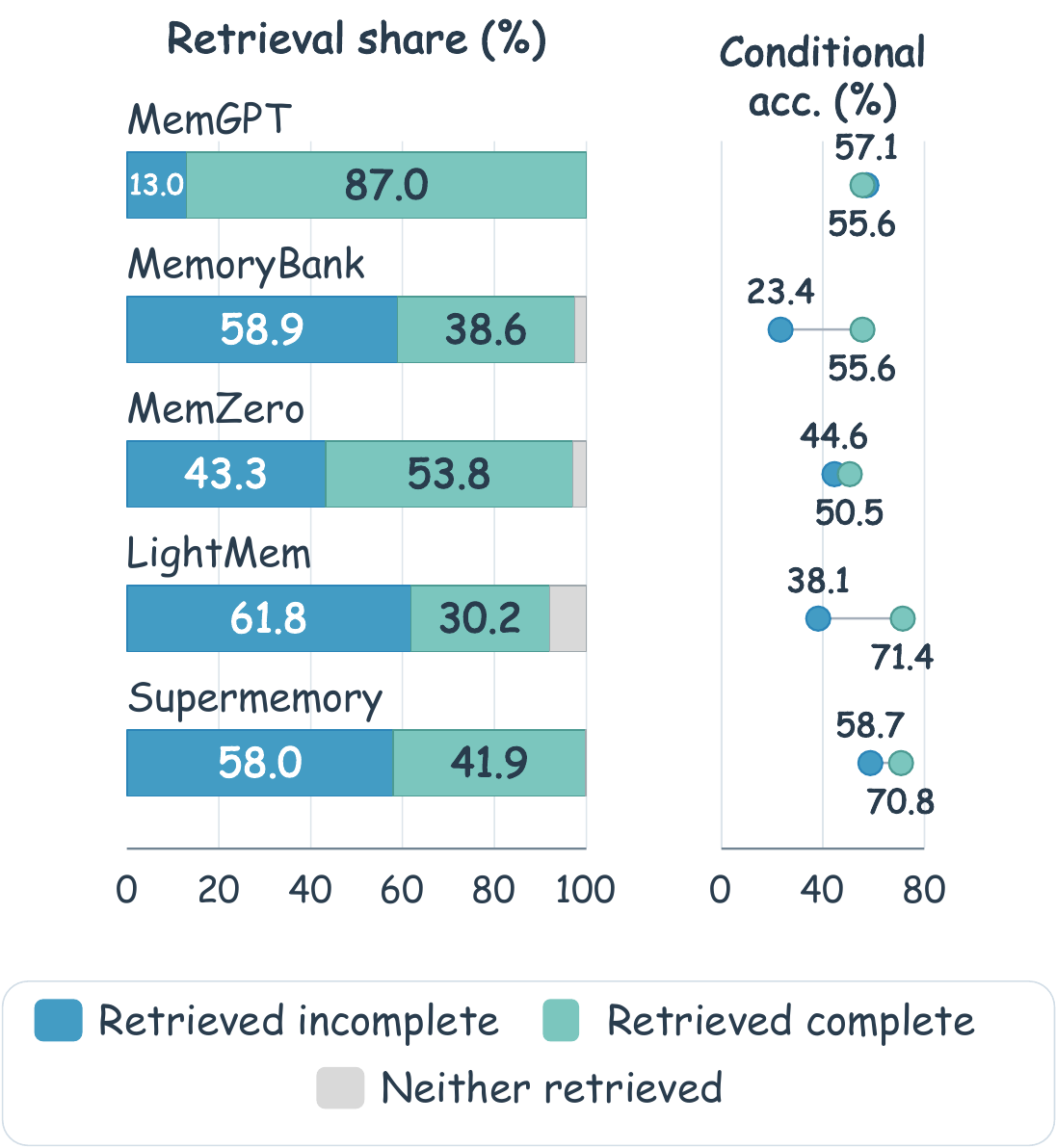}
    \caption{Retrieval completeness and conditional answer accuracy on the Valid Memory Selection subset of MemSyco-Bench across evaluated systems.}
    \label{fig:pre1}
    \vspace{-3mm}
\end{figure}

\subsection{Do Fact-Centric Systems Capture Relations among Memories?}
\label{app:prelim-relational-retrieval}

To examine whether fact-centric memory systems capture relations among memories, we evaluate five systems. As shown in Figure~\ref{fig:pre1}, MemoryBank, LightMem, and Supermemory do not retrieve all relevant information in 58.9\%, 61.8\%, and 58.0\% of cases, respectively. For four of the five systems, answer accuracy is 5.9--33.3\% higher when all required memories are retrieved than when some are missing. These findings motivate explicit modeling of memory relations, so that individual facts can be interpreted alongside the information affecting their validity and scope.

\subsection{Can Personal Agents Use Fully Retrieved Memories Appropriately?}
\label{app:prelim-memory-use}

\begin{figure}[t]
    \centering
    \includegraphics[width=\columnwidth]{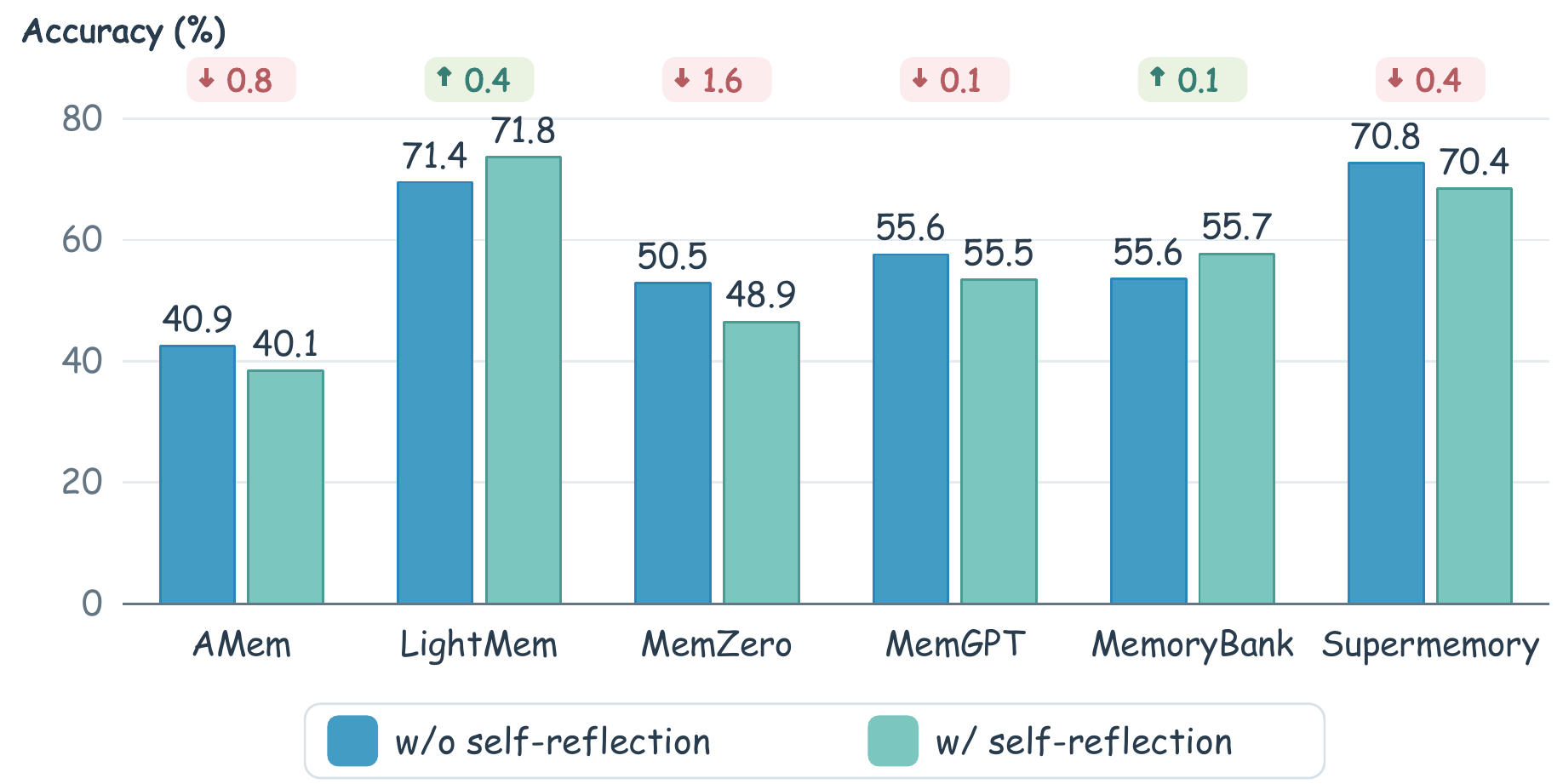}
    \caption{Effect of fixed self-reflection on answer accuracy when complete memory evidence is available.}
    \label{fig:prelim-fixed-guidance}
    \vspace{-3mm}
\end{figure}

To examine whether personal agents use memories appropriately when all relevant memories are available, we compare direct reasoning with a fixed self-reflection that asks the agent to reconsider memories before responding. As shown in Figure~\ref{fig:prelim-fixed-guidance}, self-reflection reduces accuracy for most systems, with the largest drop of 1.6\% on Mem0, and lowers average accuracy from 50.5\% to 48.9\%. These results show that access to all relevant memories does not ensure appropriate use, and the fixed reflection prompt offers no consistent improvement. This motivates learning memory-use strategies from experience to determine how memories should be applied and prioritized for each request.

\section{Method}
\label{sec:method}

\begin{figure*}[t]
    \centering
    \includegraphics[width=\textwidth]{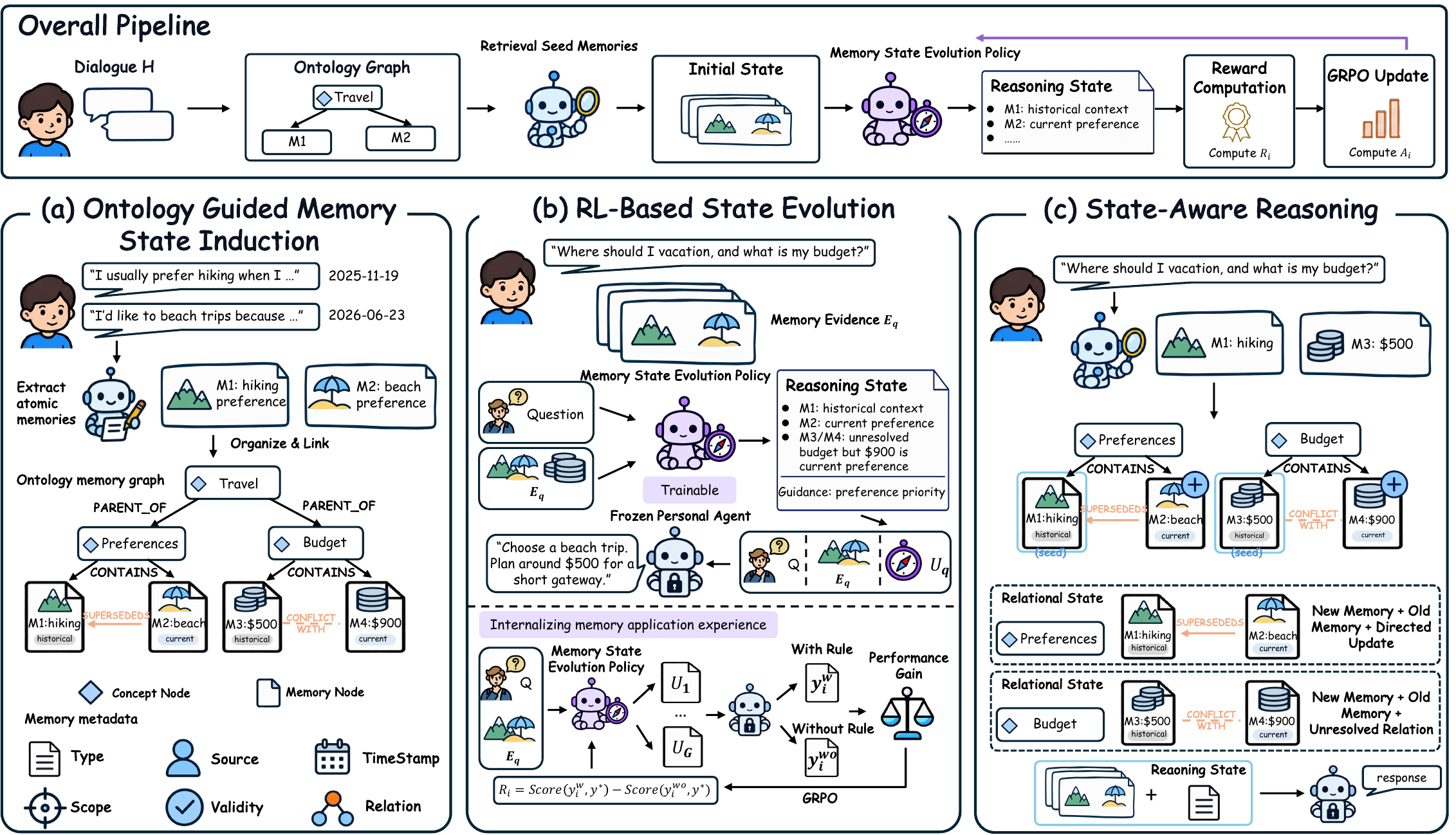}
    \caption{Overview of \sysname{} for personal agents. Ontology-guided induction transforms accumulated personal information into Persistent Facts and Relational States. RL-based state evolution learns a Memory State Evolution Policy from memory-use experience and generates a request-specific Reasoning State. State-Aware Reasoning combines the Relational States and Reasoning State to guide the personal agent's response.}
    \label{fig:method-overview}
    \vspace{-3mm}
\end{figure*}

As illustrated in Figure~\ref{fig:method-overview}, \sysname{} organizes personal memory around Persistent Facts and two complementary states. \textbf{Ontology-Guided Memory State Induction} extracts Persistent Facts from accumulated personal information and constructs Relational States that track their validity, scope, and evolving relations. \textbf{RL-Based State Evolution} learns a Memory State Evolution Policy that generates a request-specific Reasoning State from relevant memory evidence. \textbf{State-Aware Reasoning} then combines the Relational States and Reasoning State to guide the personal agent.

\subsection{System Concepts}
\label{sec:system-concepts}

Let $\mathcal{F}_t=\{f_i\}_{i=1}^{N_t}$ denote the Persistent Facts accumulated through turn $t$, and let $q_t$ denote the current request. We characterize personal memory using two complementary states; the complete formulation is provided in Appendix~\ref{app:system-concepts}.

\noindent\textbf{Definition 1 (Relational States).}
Each fact $f_i$ is associated with a Relational State:
\begin{equation}
\small
r_i^t=(f_i,a_i^t,\rho_i^t),
\qquad
\mathcal{R}_t=\{r_i^t\mid f_i\in\mathcal{F}_t\},
\end{equation}
where $a_i^t$ specifies the current validity and scope of $f_i$, and $\rho_i^t$ records how it updates, conflicts with, or constrains other facts. Relational States persist across requests and evolve as new conversations update the memory.

\noindent\textbf{Definition 2 (Reasoning State).}
For request $q_t$, the Reasoning State is inferred from the request, Persistent Facts, and their Relational States:
\begin{equation}
\small
\mathcal{Z}_{q_t}
=
\Psi(q_t,\mathcal{F}_t,\mathcal{R}_t),
\end{equation}
where $\mathcal{Z}_{q_t}$ specifies the applicability, role, and priority of memory evidence. The resulting Memory State is
$\mathcal{S}_t(q_t)=(\mathcal{R}_t,\mathcal{Z}_{q_t})$:
Relational States represent whether facts remain valid and how they are related, while the Reasoning State determines how they should be used for the current request.

\subsection{Ontology-Guided State Induction}
\label{sec:initial-memory-states}

Persistent Facts alone are insufficient to construct a memory state because their current meaning depends on their validity, scope, and relations to other facts. Therefore, we construct a personal memory ontology incrementally and induce Relational States from its evolving structure. Persistent Facts retain the original information, while their Relational States are updated as new facts are added. The resulting ontology and Relational States form the persistent personal memory.

\paragraph{Iterative ontology construction.}
Given the conversations accumulated through turn $t$, denoted by $\mathcal{H}_t$, we extract a time-ordered set of Persistent Facts:
\begin{equation}
\small
    \mathcal{F}_t
    =
    \left\{
        f_1,\ldots,f_{N_t}
    \right\}.
\end{equation}
Each $f_i$ contains a single proposition about a user preference, experience, belief, circumstance, or constraint, together with its source and time. Starting from an empty ontology $\mathcal{G}^{(0)}$, we incorporate each fact in temporal order:
\begin{equation}
\small
    \mathcal{G}^{(i)}
    =
    \operatorname{Induce}
    \left(
        \mathcal{G}^{(i-1)},
        f_i
    \right),
\end{equation}
where
\begin{equation}
\small
    \mathcal{G}^{(i)}
    =
    \left(
        \mathcal{V}_C^{(i)}
        \cup
        \mathcal{V}_F^{(i)},
        \mathcal{E}_H^{(i)}
        \cup
        \mathcal{E}_R^{(i)}
    \right).
\end{equation}
Here, $\mathcal{V}_C^{(i)}$ and $\mathcal{V}_F^{(i)}$ denote the concept and fact nodes after incorporating $f_i$. The structural edges $\mathcal{E}_H^{(i)}$ use \texttt{PARENT\_OF} to organize concept nodes hierarchically and \texttt{CONTAINS} to attach each fact to its most specific concept. The \texttt{CREATE}, \texttt{MERGE}, and \texttt{SPLIT} operations revise the concept hierarchy as new facts are added. The relation edges $\mathcal{E}_R^{(i)}$ use \texttt{SUPERSEDES} to represent updates and \texttt{CONFLICTS\_WITH} to preserve unresolved conflicts.

\paragraph{Ontology-induced Relational States.}
The Relation State of each fact is derived from its current attributes and ontology relations. After incorporating the first $i$ facts, we define
\begin{equation}
\small
    \mathcal{R}^{(i)}
    =
    \left\{
        r_f^{(i)}
        =
        \left(
            f,
            a_f^{(i)},
            \rho_f^{(i)}
        \right)
        \,\middle|\,
        f\in\mathcal{V}_F^{(i)}
    \right\},
    \label{eq:ontology-relation-states}
\end{equation}
where $a_f^{(i)}$ describes the current validity and scope of $f$, and $\rho_f^{(i)}$ contains its concept attachment and relations to other facts. Later facts may update these attributes and relations without overwriting $f$. After all facts have been incorporated, we obtain
\begin{equation}
\small
    \mathcal{R}_t
    =
    \mathcal{R}^{(N_t)}.
\end{equation}
The Persistent Facts $\mathcal{F}_t$ and Relational States $\mathcal{R}_t$ are retained as personal memory and provide the persistent inputs for inferring a Reasoning State for each new user request.

\subsection{RL-Based State Evolution}
\label{sec:dynamic-memory-states}

Relational States capture how facts currently hold and relate, but do not specify how a personal agent should use them for a particular request. We therefore treat state evolution as the internalization of reusable memory-use experience and instantiate the state inference function $\Psi$ with a Memory State Evolution Policy $\pi_\theta$. Given request $q$, relevant Persistent Facts $\mathcal{F}_q$, and their Relational States $\mathcal{R}_q$, the policy generates a Reasoning State:
\begin{equation}
\small
    \mathcal{Z}_q
    \sim
    \pi_\theta
    \left(
        \cdot
        \mid
        q,
        \mathcal{F}_q,
        \mathcal{R}_q
    \right).
\end{equation}
We train $\pi_\theta$ by repeatedly generating and evaluating candidate Reasoning States, allowing the policy to learn reusable patterns of memory use.

\paragraph{Evaluating candidate Reasoning States.}
A Reasoning State is useful only if it helps the personal agent use the given memory evidence correctly. Let
\begin{equation}
\small
    \mathcal{M}_q
    =
    \left(
        \mathcal{F}_q,
        \mathcal{R}_q
    \right)
\end{equation}
denote the relevant memory evidence for request $q$. Given a candidate Reasoning State $\mathcal{Z}_q^{(i)}$, we evaluate its effect on a frozen personal agent $\mathcal{A}$ relative to a baseline that receives the same memory evidence without a Reasoning State:
\begin{equation}
\small
\begin{aligned}
    R_i ={}&
    \operatorname{Score}
    \left(
        \mathcal{A}
        \left(
            q,
            \mathcal{M}_q,
            \mathcal{Z}_q^{(i)}
        \right),
        y^*
    \right) \\
    &-
    \operatorname{Score}
    \left(
        \mathcal{A}
        \left(
            q,
            \mathcal{M}_q
        \right),
        y^*
    \right),
\end{aligned}
\label{eq:relative-gain}
\end{equation}
where $y^*$ is the reference answer and $\operatorname{Score}$ is the task-specific evaluator. A positive reward indicates that the candidate improves how the personal agent uses the fixed memory evidence, whereas a negative reward penalizes candidates that distort or weaken that evidence.

\paragraph{Learning the state evolution policy.}
We optimize $\pi_\theta$ using GRPO~\citep{shao2024deepseekmath}. For each training instance $(q,\mathcal{M}_q,y^*)$, we sample $G$ candidate Reasoning States from the previous policy:
\begin{equation}
\small
    \mathcal{Z}_q^{(i)}
    \sim
    \pi_{\theta_{\mathrm{old}}}
    \left(
        \cdot
        \mid
        q,
        \mathcal{F}_q,
        \mathcal{R}_q
    \right),
    \qquad i=1,\ldots,G.
\end{equation}
Their relative rewards are normalized within the group to obtain advantages $\widehat{A}_i$, which guide the policy update:
\begin{equation}
\small
\begin{aligned}
    \mathcal{L}_{\mathrm{GRPO}}(\theta)
    ={}&
    -\frac{1}{G}
    \sum_{i=1}^{G}
    \frac{1}{T_i}
    \sum_{j=1}^{T_i}
    \min
    \Bigl(
        \rho_{i,j}\widehat{A}_i, \\
    &\operatorname{clip}
        \left(
            \rho_{i,j},
            1-\epsilon,
            1+\epsilon
        \right)
        \widehat{A}_i
    \Bigr) \\
    &+
    \beta_{\mathrm{KL}}
    \mathcal{D}_{\mathrm{KL}}
    \left(
        \pi_\theta
        \,\Vert\,
        \pi_{\mathrm{ref}}
    \right).
\end{aligned}
\label{eq:grpo}
\end{equation}
Here, $T_i$ is the length of candidate $\mathcal{Z}_q^{(i)}$, $\rho_{i,j}$ is the token-level probability ratio between $\pi_\theta$ and $\pi_{\theta_{\mathrm{old}}}$, $\epsilon$ is the clipping threshold, and $\beta_{\mathrm{KL}}$ controls regularization toward the reference policy $\pi_{\mathrm{ref}}$. Through repeated optimization, $\pi_\theta$ internalizes memory-use patterns that generalize to new facts and requests. Training details are provided in Appendix~\ref{app:hint-interface}.

% \paragraph{Forming memory reasoning states.}
% After training, the policy uses the current request and its related initial states to produce a dynamic use state:
% \begin{equation}
%     \mathcal{Z}_q
%     =
%     \pi_\theta(q,\mathcal{E}_q),
%     \qquad
%     \mathcal{S}_q^{\mathrm{dyn}}
%     =
%     (\mathcal{E}_q,\mathcal{Z}_q).
% \label{eq:memory-state}
% \end{equation}
% The learned policy stores experience from repeated memory use in its parameters. It can therefore form task-specific states for new requests without changing the underlying factual records. The next stage uses $\mathcal{S}_q^{\mathrm{dyn}}$ to support state-aware reasoning.

\subsection{State-Aware Reasoning}
\label{sec:state-aware-reasoning}

At inference, State-Aware Reasoning retrieves the Persistent Facts and Relational States required by the current request, generates a Reasoning State, and grounds the personal agent in both states.

\begingroup
% Packages: array, booktabs, multirow, graphicx, colortbl/xcolor.
% Keep the manuscript's existing \sysname and citation commands.
\def\TBHead#1{{\fontsize{6.1}{7.2}\selectfont\shortstack[c]{#1}}}
\def\TBMethod#1{{\fontsize{6.4}{8.6}\selectfont #1}}
\begin{table*}[!t]
\centering
\fontsize{7.2}{8.6}\selectfont
\renewcommand{\arraystretch}{1.12}
\setlength{\tabcolsep}{1.25pt}
\def\TBMethodWidth{4cm}
\edef\TBMetricWidth{\the\dimexpr(\textwidth-\TBMethodWidth-30\tabcolsep)/15\relax}
\begin{tabular}{@{}>{\raggedright\arraybackslash}m{\TBMethodWidth}*{15}{>{\centering\arraybackslash}m{\TBMetricWidth}}@{}}
\toprule
\TBHead{\textbf{Memory}\\\textbf{System}}
& \multicolumn{5}{c}{\textbf{MemSyco-Bench}}
& \multicolumn{5}{c}{\textbf{MemTrapBench}}
& \multicolumn{2}{c}{\TBHead{\textbf{LoCoMo-}\\\textbf{Refined}}}
& \multicolumn{2}{c}{\textbf{LoCoMo}}
& \TBHead{\textbf{Overall}\\\textbf{Avg.} $\uparrow$}
\\
\cmidrule(lr){2-6}\cmidrule(lr){7-11}\cmidrule(lr){12-13}\cmidrule(lr){14-15}
& \TBHead{WTU\\Acc. $\uparrow$}
& \TBHead{WTU\\Syco. $\downarrow$}
& \TBHead{HTU\\Acc. $\uparrow$}
& \TBHead{HTU\\Use $\uparrow$}
& \TBHead{Avg. $\uparrow$}
& \TBHead{Safety\\Score $\uparrow$}
& \TBHead{CogBias\\Acc. $\uparrow$}
& \TBHead{Bound.\\Acc. $\uparrow$}
& \TBHead{Trauma\\Acc. $\uparrow$}
& \TBHead{Avg. $\uparrow$}
& \TBHead{Judge\\Score $\uparrow$}
& \TBHead{Token\\F1 $\uparrow$}
& \TBHead{Judge\\Score $\uparrow$}
& \TBHead{Token\\F1 $\uparrow$}
& \\
\specialrule{0.5pt}{2pt}{3pt}
% =========================================================
% Qwen3-8B
% =========================================================

\rowcolor[HTML]{EFEFEF}
\multicolumn{16}{c}{\textbf{Qwen3-8B}} \\
\midrule

\TBMethod{NaiveRAG~\citep{lewis2020retrieval}}
& 36.3
& 55.3
& 51.6
& 64.0
& 42.4
& 60.2
& 63.9
& 77.0
& \underline{93.4}
& 73.6
& 39.0
& 38.4
& 34.4
& 35.8
& 47.4
\\

\TBMethod{MemGPT~\citep{packer2023memgpt}}
& 29.0
& 66.7
& 54.0
& 67.0
& 39.0
& 62.2
& 66.0
& 64.5
& 52.4
& 61.3
& 45.7
& 40.0
& 36.8
& 37.0
& 45.7
\\

\TBMethod{MemoryBank~\citep{zhong2023memorybank}}
& 37.5
& 57.6
& 54.4
& 67.1
& 44.2
& 64.9
& 63.7
& 72.1
& 82.3
& 70.7
& 44.9
& 44.0
& 39.9
& 41.0
& 49.9
\\

\TBMethod{Mem0~\citep{chhikara2025mem0}}
& 25.9
& \underline{47.8}
& 52.0
& 63.8
& 36.3
& 56.8
& \underline{69.3}
& 81.9
& 85.5
& 73.4
& 25.7
& 26.9
& 21.4
& 23.8
& 39.2
\\

% A-MEM: user-supplied scores; 0--1 scores converted to percentages.
\TBMethod{A-MEM~\citep{xu2025amem}}
& \underline{42.8}
& 49.1
& 49.2
& 61.2
& \underline{45.4}
& 64.2
& 65.0
& 77.6
& 91.8
& 74.6
& 41.8
& 39.1
& 36.4
& 37.1
& 49.6
\\

\TBMethod{LightMem~\citep{wu2025lightmem}}
& 21.0
& 49.6
& 46.1
& 58.7
& 31.0
& 68.6
& \textbf{69.9}
& \underline{88.6}
& 91.6
& \underline{79.7}
& 25.3
& 27.2
& 22.1
& 24.0
& 39.5
\\

\TBMethod{Supermemory~\citep{supermemory2026}}
& 23.3
& 70.5
& \underline{63.7}
& \underline{74.2}
& 39.5
& 67.5
& 65.4
& 79.6
& 61.7
& 68.6
& 16.6
& 17.5
& 15.7
& 15.8
& 35.1
\\

\TBMethod{SimpleMem~\citep{liu2026simplemem}}
& 33.4
& 58.6
& 44.1
& 52.6
& 37.7
& 70.9
& 68.3
& 84.6
& \textbf{95.8}
& \textbf{79.9}
& \underline{51.3}
& \underline{46.6}
& \underline{45.3}
& \textbf{46.4}
& \underline{53.6}
\\

\TBMethod{PGMem~\citep{choi2026pgmem}}
& 33.2
& 62.8
& 54.5
& 69.1
& 41.8
& \underline{71.0}
& 66.4
& 85.1
& 75.6
& 74.5
& 28.5
& 26.9
& 25.4
& 25.2
& 42.6
\\

\rowcolor{blue!8}
\TBMethod{\textbf{\sysname{} (Ours)}}
& \textbf{47.7}
& \textbf{38.0}
& \textbf{72.4}
& \textbf{81.3}
& \textbf{57.6}
& \textbf{73.8}
& 62.3
& \textbf{88.8}
& 91.4
& 79.1
& \textbf{57.4}
& \textbf{47.3}
& \textbf{45.5}
& \underline{41.6}
& \textbf{59.9}
\\

% =========================================================
% GPT-4o-mini
% =========================================================

\midrule
\rowcolor[HTML]{EFEFEF}
\multicolumn{16}{c}{\textbf{GPT-4o-mini}} \\
\midrule

\TBMethod{NaiveRAG~\citep{lewis2020retrieval}}
& 48.3
& 42.8
& 49.2
& 61.1
& 48.7
& 61.7
& 58.6
& 65.0
& \textbf{97.4}
& 70.7
& 43.2
& 39.1
& 34.4
& 36.9
& 49.3
\\

\TBMethod{MemGPT~\citep{packer2023memgpt}}
& 40.8
& 48.0
& \underline{56.2}
& 66.0
& 46.9
& 66.4
& 60.7
& 69.9
& 62.3
& 64.8
& 49.1
& 39.5
& 40.2
& 39.1
& 50.3
\\

\TBMethod{MemoryBank~\citep{zhong2023memorybank}}
& 48.9
& 42.2
& 48.3
& 62.3
& 48.7
& 66.4
& 59.8
& 73.4
& 90.5
& 72.5
& 52.8
& 46.6
& 42.7
& \underline{45.3}
& \underline{54.2}
\\

\TBMethod{Mem0~\citep{chhikara2025mem0}}
& 32.7
& 38.4
& 50.3
& 63.3
& 39.7
& 56.4
& 61.7
& 74.1
& 93.5
& 71.4
& 28.9
& 27.2
& 21.6
& 25.2
& 40.4
\\

% A-MEM: user-supplied scores; rounded to one decimal using half-up.
\TBMethod{A-MEM~\citep{xu2025amem}}
& \textbf{54.5}
& \underline{35.7}
& 50.9
& 63.1
& \underline{53.0}
& 63.1
& 59.6
& 69.3
& 95.8
& 71.9
& 49.1
& 42.7
& 39.4
& 41.9
& 53.4
\\

\TBMethod{LightMem~\citep{wu2025lightmem}}
& 25.0
& 42.9
& 46.9
& 60.4
& 33.8
& 70.2
& 63.8
& 80.8
& 95.2
& \underline{77.5}
& 24.2
& 26.9
& 21.8
& 25.1
& 39.3
\\

\TBMethod{Supermemory~\citep{supermemory2026}}
& 35.0
& 56.3
& 52.9
& 65.9
& 42.2
& 67.4
& 58.8
& 74.2
& 76.4
& 69.2
& 15.8
& 16.0
& 14.9
& 14.8
& 35.5
\\

\TBMethod{SimpleMem~\citep{liu2026simplemem}}
& 25.0
& 63.2
& 47.6
& 60.7
& 34.0
& 69.7
& \underline{64.6}
& 72.7
& \underline{97.3}
& 76.1
& \underline{53.8}
& \underline{48.0}
& \textbf{47.2}
& \textbf{48.4}
& 52.8
\\

\TBMethod{PGMem~\citep{choi2026pgmem}}
& 39.6
& 54.4
& 53.7
& \underline{66.3}
& 45.2
& \underline{73.9}
& 60.9
& \underline{81.1}
& 83.1
& 74.7
& 19.8
& 18.9
& 16.7
& 17.6
& 39.1
\\

\rowcolor{blue!8}
\TBMethod{\textbf{\sysname{} (Ours)}}
& \underline{53.1}
& \textbf{31.9}
& \textbf{67.6}
& \textbf{80.1}
& \textbf{58.9}
& \textbf{73.9}
& \textbf{73.9}
& \textbf{88.8}
& 90.4
& \textbf{78.7}
& \textbf{60.8}
& \textbf{48.5}
& \underline{46.4}
& 45.2
& \textbf{61.2}
\\

\bottomrule
\end{tabular}
\caption{Main results on four benchmarks with Qwen3-8B and GPT-4o-mini. $\uparrow$ and $\downarrow$ denote metrics to maximize and minimize, respectively. \textbf{Bold} and \underline{underlined} values denote the best and second best results for each backbone. WTU/HTU: When/How to Use memory; Syco.: sycophancy rate; Use: correct use.}
\label{tab:main-results}
\vspace{-3mm}
\end{table*}
\endgroup

Given a request $q$, we retrieve $K$ semantically similar Persistent Facts as seeds. Because a seed may be updated, constrained, or contradicted by other facts, we expand it by traversing the stored memory relations in both directions:
\begin{equation}
\small
    \mathcal{F}_q
    =
    \bigcup_{
        f\in
        \operatorname{Retrieve}_{\mathrm{sim}}
        (q,\mathcal{F}_t,K)
    }
    \operatorname{Traverse}
    \left(
        f,
        \mathcal{R}_t
    \right).
    \label{eq:relevant-memory-evidence}
\end{equation}
Here, $\operatorname{Traverse}(f,\mathcal{R}_t)$ includes $f$ and all facts connected to it through memory relations. Facts reached from multiple seeds are included only once. We denote the Relational States corresponding to $\mathcal{F}_q$ as $\widetilde{\mathcal{R}}_q$.

The learned policy generates a Reasoning State from the request, retrieved facts, and their Relational States. The personal agent then produces its response using both states:
\begin{equation}
\small
\begin{aligned}
    \mathcal{Z}_q
    &=
    \pi_\theta
    \left(
        q,
        \mathcal{F}_q,
        \widetilde{\mathcal{R}}_q
    \right), \\
    y
    &=
    \mathcal{A}
    \left(
        q,
        \mathcal{F}_q,
        \widetilde{\mathcal{R}}_q,
        \mathcal{Z}_q
    \right).
\end{aligned}
\label{eq:state-aware-answer}
\end{equation}
Together, relation traversal preserves the context needed to interpret the retrieved facts, while the Reasoning State specifies how the personal agent should use them for the current request.

% Q1--Q3 experiment draft for MemBridge.
% Required packages: booktabs, array, multirow, makecell, graphicx, xcolor[table].
% Chinese companion text is retained as comments for revision.
\section{Experiment}

% Required packages:
% \usepackage{booktabs}
% \usepackage{multirow}
% \usepackage{graphicx}
% \usepackage[table]{xcolor}

% 本节围绕五个问题评估 MemBridge：主实验、两个缺口的解决情况、指导策略的泛化性、消融实验，以及附录中的案例与效率分析。
% \paragraph{Experimental Questions.}
% In this section, we evaluate MemBridge by answering the following questions.
% \textbf{Q1 (Main Results):} How does MemBridge compare with existing long-term memory systems across benchmarks and response models?
% \textbf{Q2 (Bridging the Two Gaps):} How effectively does MemBridge address the Evidence Connection Gap and the Memory Use Gap?
% \textbf{Q3 (Guidance Generalization):} How well does the learned guidance policy generalize across memory-use categories, memory systems, and benchmarks?
% \textbf{Q4 (Ablation Study):} How does each component contribute to overall performance?
% We further provide case studies and efficiency analysis (\textbf{Q5}) in the appendix.

In this section, we evaluate how \sysname{} constructs and applies memory states through the following four research questions. \textbf{Q1 (Main Results):} How does \sysname{} compare with existing memory systems? \textbf{Q2 (Memory State Effectiveness):} How effectively do the relational state and reasoning state support reliable memory use? \textbf{Q3 (Reasoning State Generalization):} How well does reasoning state generation transfer across tasks and memory systems? \textbf{Q4 (Ablation Study):} How does each component affect overall performance? Appendix~\ref{app:additional-experiments} provides a case study and efficiency analysis.

\begin{figure*}[t]
    \centering
    \includegraphics[width=\textwidth]{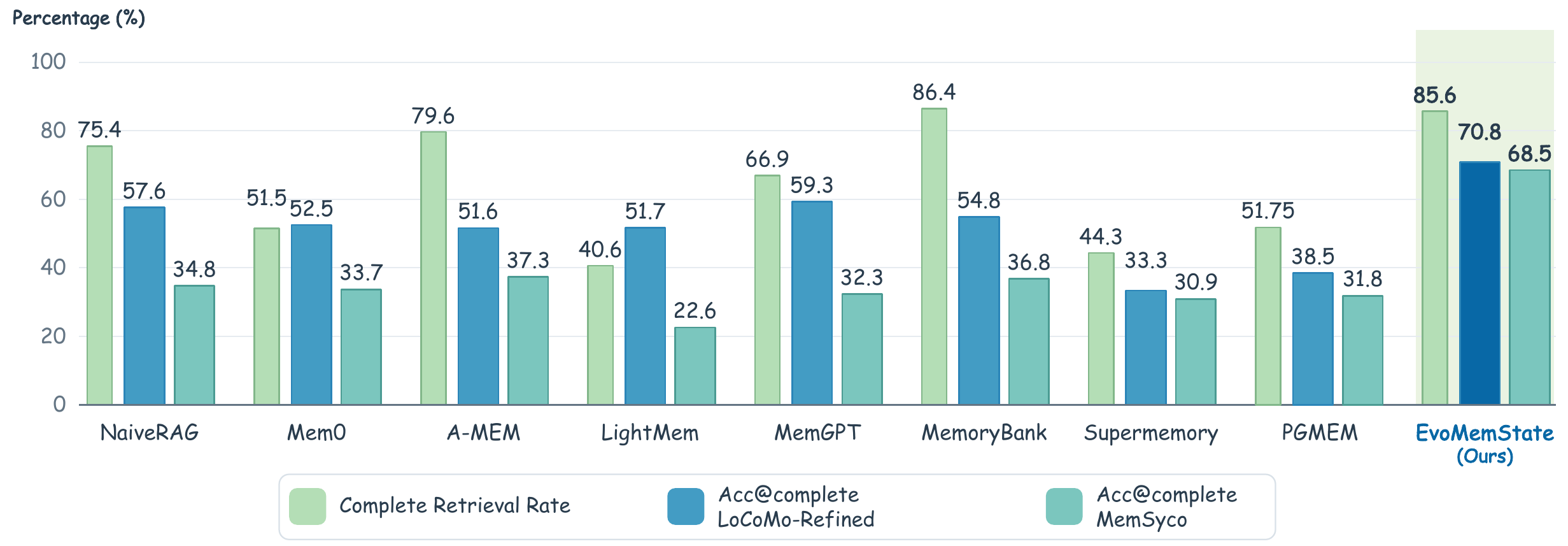}
    \caption{Complete retrieval rate and accuracy on questions with complete evidence using Qwen3-8B. Average complete retrieval rate reports the percentage of questions with all required evidence retrieved. Acc@complete reports accuracy on LoCoMo-Refined and MemSyco-Bench for questions with all required evidence retrieved.}
    \label{fig:memory_use}
    \vspace{-3mm}
\end{figure*}

\subsection{Experimental Setup}
\label{sec:experimental-setup}

% \paragraph{Training Data Construction.}
% We construct approximately 60K question-memory instances from FEVER~\citep{thorne2018fever}, SciFact~\citep{wadden2020scifact}, and PersonaMem~\citep{jiang2025personamem} to train the Hint Policy. FEVER and SciFact provide factual questions with supporting evidence, to which we add conflicting memories about user beliefs, identity, and perceived expertise. PersonaMem provides preference scenarios covering current preferences, earlier preferences, and updates. We add unrelated memories to both types of instances so that the policy must determine which records apply to the question. Details are given in Appendix~\ref{app:training-data-construction}.

\paragraph{Datasets.}
We evaluate on four benchmarks covering memory use in personal agents and long term conversational memory. MemSyco-Bench~\citep{xiang2026memsyco} and MemTrapBench~\citep{fang2026memtrap} assess personal agents' use of historical memories, including preference updates, evidence priority, and effects on reasoning. LoCoMo~\citep{maharana2024locomo} and LoCoMo-Refined\footnote{\url{https://github.com/mem-eval-suite/LoCoMo_refined}} evaluate recall and reasoning across dialogue sessions. Dataset details appear in Appendix~\ref{app:benchmark-datasets}.

\paragraph{Baselines.}
We compare \sysname{} with nine baselines covering retrieval, persistent memory, and structured memory organization: NaiveRAG~\citep{lewis2020retrieval}, MemGPT~\citep{packer2023memgpt}, MemoryBank~\citep{zhong2023memorybank}, Mem0~\citep{chhikara2025mem0}, LightMem~\citep{wu2025lightmem}, A-MEM~\citep{xu2025amem}, Supermemory~\citep{supermemory2026}, SimpleMem~\citep{liu2026simplemem}, and PGMem~\citep{choi2026pgmem}. Details are provided in Appendix~\ref{app:baseline-details}.

\paragraph{Evaluation Metrics.}
We evaluate answer quality, memory use, and retrieval completeness. Answer quality is measured by accuracy, F1, and MemTrapBench judge scores. Following MemSyco-Bench~\citep{xiang2026memsyco}, we use Syco. Rate to measure inappropriate reliance on historical memories and Correct Use to measure valid personalization. Retrieval is evaluated by Complete Retrieval Rate and answer accuracy on complete-retrieval cases (Acc@complete). Full definitions are provided in Appendix~\ref{app:evaluation-metrics}.

\paragraph{Implementation Details.}
We use DeepSeek-V4-Flash for offline memory construction and Qwen3-8B~\citep{yang2025qwen3} and GPT-4o-mini as online response models. By default, the Memory State Evolution Policy is initialized from Qwen3-4B-Instruct-2507~\citep{yang2025qwen3} and trained with GRPO~\citep{shao2024deepseekmath}. Dataset construction, training details, and prompts are provided in Appendices~\ref{app:benchmark-datasets}, \ref{app:hint-interface}, and~\ref{app:prompt-set}.

\subsection{Main Results (Q1)}

We compare \sysname{} with memory systems using Qwen3-8B and GPT-4o-mini. Table~\ref{tab:main-results} reports results on MemSyco-Bench and MemTrapBench for memory use, and on LoCoMo and LoCoMo-Refined for long-term memory reasoning.

\noindent\textbf{Obs.1. \sysname{} improves when and how memory is used.}
On MemSyco-Bench, \sysname{} achieves accuracies of 57.6\% and 58.9\% with Qwen3-8B and GPT-4o-mini, outperforming the strongest baselines by 12.2\% and 5.9\%, respectively. It achieves the best When to Use and How to Use accuracy, the lowest Syco. Rate, and the highest Correct Use with Qwen3-8B. These results show more reliable decisions about whether and how history should affect the current response.

\noindent\textbf{Obs.2. \sysname{} supports long-term recall and task-appropriate reasoning.}
On LoCoMo-Refined, \sysname{} achieves the highest LLM-judge and token F1 scores with both models, outperforming the strongest baseline in judge score by 6.1\% and 7.0\%. It also achieves the best Safety and Task Boundary scores on MemTrapBench, showing that factual records support recall while reasoning states prevent irrelevant history and prior task rules from overriding the current request.

\subsection{Memory-State Effectiveness (Q2)}

To assess the two memory states, we examine whether the Relational State assembles complete evidence and whether the Reasoning State uses it correctly. Figure~\ref{fig:memory_use} reports Complete Retrieval Rate and Acc@complete on LoCoMo-Refined and MemSyco-Bench with Qwen3-8B; the metrics are considered jointly because complete-evidence subsets differ across systems.

\noindent\textbf{Obs.3. The Relational State connects related memories into reliable evidence.}
\sysname{} achieves a complete retrieval rate of 85.6\%, comparable to the best baseline. Guided by the personal memory ontology, its Relational State preserves each fact's current attributes together with related updates, conflicts, and constraints. This organization allows related memories to provide complete evidence rather than remain independent records.

\noindent\textbf{Obs.4. The Reasoning State supports reliable use of complete evidence.}
Acc\@complete reaches 70.8\% on LoCoMo-Refined and 68.5\% on MemSyco-Bench, exceeding the strongest baselines by 11.5\% and 31.2\%, respectively. These gains show that evidence completeness alone does not ensure correct use. By internalizing memory-use experience, the learned policy generates a Reasoning State that makes the applicability, role, and priority of each memory explicit, helping the response model apply the assembled evidence appropriately and consistently.

\subsection{Reasoning-State Generalization (Q3)}

\begin{figure}[t]
    \centering
    \includegraphics[width=0.5\textwidth]{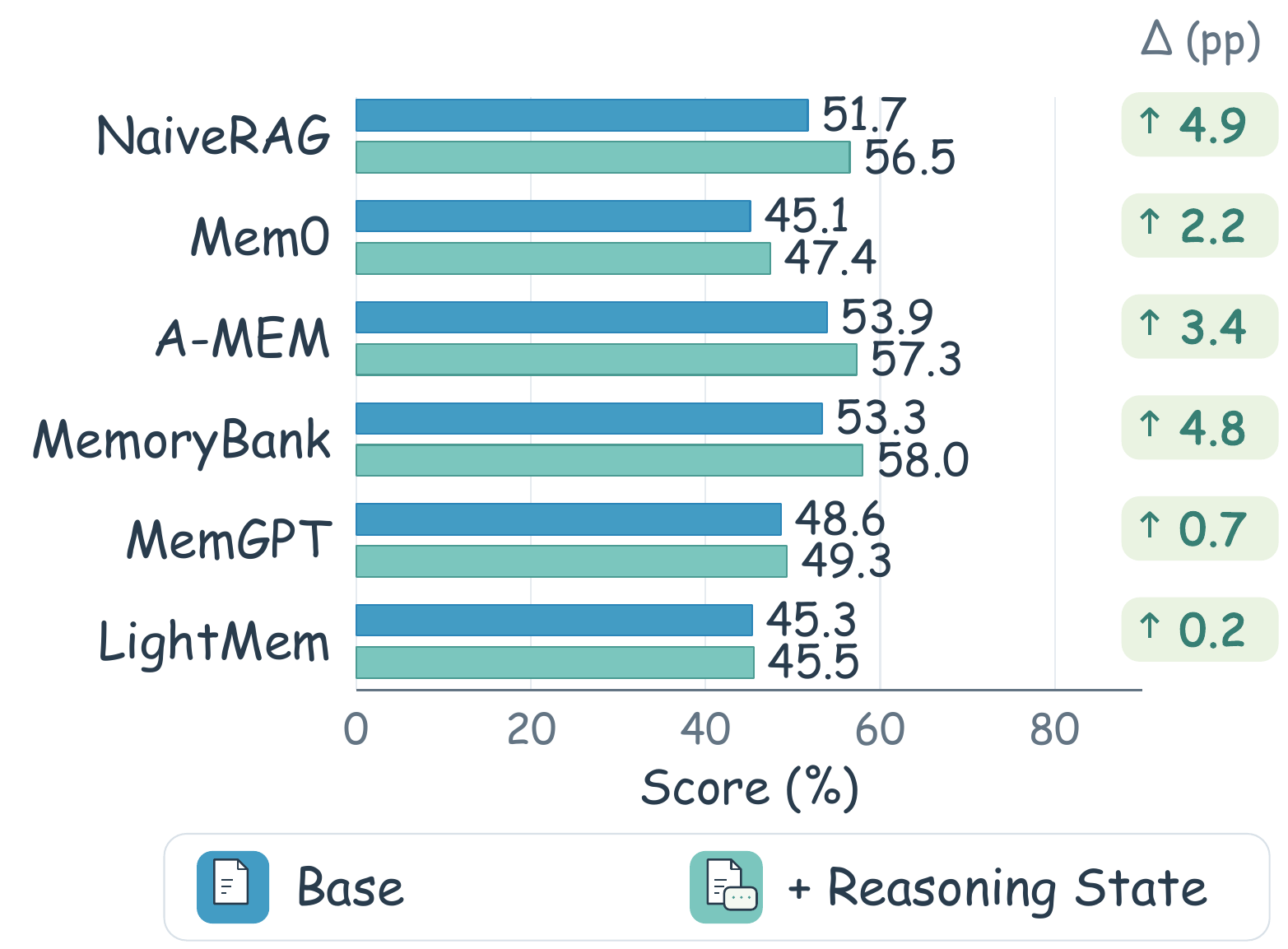}
    \caption{Cross-system transfer of the Memory State Evolution Policy, averaged over MemSyco-Bench, MemTrapBench, and LoCoMo-Refined. Base and +Reasoning State denote results without and with generated reasoning states; $\Delta$ is the gain.}
    \label{fig:hint_bridge_comparison}
    \vspace{-3mm}
\end{figure}

To evaluate whether reasoning-state generation generalizes across tasks and memory systems, we pair the Memory State Evolution Policy with different memory systems on the benchmarks. Figure~\ref{fig:hint_bridge_comparison} reports cross-system performance changes, while Appendix~\ref{app:memory-use-categories} compares alternative reasoning-state strategies across memory-use categories.

\noindent\textbf{Obs.5. Reasoning states transfer across memory systems.}
All six baselines improve when augmented with generated reasoning states, with overall gains ranging from 0.2\% to 4.9\%. NaiveRAG and MemoryBank benefit most, improving by 4.9\% and 4.8\%, followed by A-MEM at 3.4\%. These results show that the policy can produce useful reasoning states from evidence supplied by different memory systems.

% Suggested appendix caption for Table~\ref{tab:memsyco_five_categories}:
% Category-level comparison of memory-use strategies on the five MemSyco tasks.
% +Guidance denotes the learned guidance policy. All values are percentages;
% bold indicates the best result in each column, and Avg. is the unweighted mean.

\begin{table}[t]
\centering
\scriptsize
\setlength{\tabcolsep}{2.6pt}
\renewcommand{\arraystretch}{1.12}

\resizebox{\columnwidth}{!}{%
\begin{tabular}{lccccc}
\toprule
\textbf{Method}
& \shortstack{\textbf{MemSyco}\\Avg.}
& \shortstack{\textbf{LoCoMo-R}\\LLM-J.}
& \shortstack{\textbf{LoCoMo}\\LLM-J.}
& \shortstack{\textbf{MemTrap}\\Avg.}
& \textbf{Overall} \\
\midrule

\rowcolor{blue!8}
\textbf{\sysname{} (Ours)}
& \textbf{57.6}
& \textbf{57.4}
& \textbf{45.5}
& \textbf{79.1}
& \textbf{59.9} \\

\quad w/o Relational State
& 48.7
& 54.9
& 45.2
& 72.7
& 55.4 \\

\quad w/o Reasoning State
& 49.6
& 55.8
& 42.5
& 64.1
& 53.0 \\

\quad w/o RL-Based State Evolution
& 49.7
& 55.4
& 45.0
& 62.7
& 53.2 \\

\quad w/ RFT
& 50.3
& 55.6
& 45.2
& 70.0
& 55.3 \\

\bottomrule
\end{tabular}%
}
\caption{Ablation results for \sysname{} with Qwen3-8B. The variants remove Relational State, Reasoning State or RL-Based State Evolution, or replace GRPO with rejection-sampling fine-tuning (RFT).}
\label{tab:component_ablation}
\vspace{-3mm}
\end{table}

\subsection{Ablation Study (Q4)}
\label{sec:ablation-study}

To examine the contribution of state composition and learned memory use (Q4), Table~\ref{tab:component_ablation} compares four variants with Qwen3-8B. \textbf{w/o Relational State} uses retrieved seed records without incorporating related updates and conflicts into the state evidence. \textbf{w/o Reasoning State} provides the personal agent with state evidence but omits the policy-generated reasoning state. \textbf{w/o RL-Based State Evolution} retains reasoning-state generation but uses the policy before RL training. \textbf{w/ RFT} trains Qwen3-4B via rejection-sampling fine-tuning on the same training data.

\noindent\textbf{Obs.6. Composing related records strengthens state evidence.} Removing relational state reduces MemSyco-Bench and the MemTrapBench score by 8.9\% and 6.4\%, respectively. The losses on LoCoMo-Refined and LoCoMo are smaller, at 2.5\% and 0.3\%. These larger losses on memory-use tasks suggest that related updates and conflicts are particularly valuable for interpreting retrieved records before applying them to the response.

\noindent\textbf{Obs.7. RL-based state evolution improves reasoning states.} Removing dynamic rules reduces MemSyco-Bench accuracy and the MemTrapBench score by 8.0\% and 15.0\%, respectively. Retaining rule generation without RL yields similar losses of 7.9\% and 16.4\%. Thus, generating rules alone provides little benefit on these benchmarks. Even on the same 60K data, GRPO beats RFT by 4.6\% overall (59.9\% vs.\ 55.3\%) with shorter, stabler outputs (94.7 vs.\ 143.2 tokens). These results support learning from memory application experience: answer-quality feedback helps the policy determine memory applicability, role, and priority, expressed as dynamic rules for each new user request.

\section{Related Work}
\subsection{Personal Agents and User Modeling}

Personal agents use persistent user information to adapt across interactions~\citep{li2026survey}. LaMP~\citep{salemi2024lamp} retrieves relevant profile items for personalized generation, while HELLO and related methods maintain and update user representations over time~\citep{li2025hello,tan2025prospect,kang2025memory}. PGMem~\citep{choi2026pgmem} links persona information to events that support or revise it. Recent work further extends personalization to Web actions and agent training~\citep{cai2025large,sun2025training}. Together, these studies cover persistent personalization across generation, user representation, and agent behavior.

\subsection{Long-Term Memory Systems}

Long-term memory systems preserve information beyond the active context for later use~\cite{ning2026memadapter,xiao2026zero}. MemGPT~\citep{packer2023memgpt} manages information across storage tiers, MemoryBank~\citep{zhong2023memorybank} maintains personal records, LightMem~\citep{wu2025lightmem} uses compression and selective retrieval, and Mem0~\citep{chhikara2025mem0} extracts and updates facts. A-MEM~\citep{xu2025amem}, Supermemory~\citep{supermemory2026}, and SimpleMem~\citep{liu2026simplemem} further organize memories through links, temporal structure~\citep{rasmussen2025zep}, and update mechanisms.

\section{Conclusion}
\label{sec:discussion}

In this paper, we presented \sysname{}, a state-centric long-term memory framework for personal agents that constructs coherent Relation States from multi-source memories and learns from continual memory use to generate Reasoning States. Experiments across memory-use and dialogue-recall benchmarks show consistent improvements, particularly under preference updates and conflicting evidence, demonstrating that reliable personal memory requires both coherent state construction and use-driven state evolution.

\section*{Limitation}
Our study focuses on text-based interaction histories. Real-world personal memory systems must also handle multimodal information, including documents, images, and other user artifacts. Extending our memory structure to represent, connect, and apply information across modalities remains an important direction for future work.

\section*{Use of Large Language Models}

During manuscript preparation, we used large language models to refine the English text, correct grammatical errors, and improve clarity and readability. We also used them to assist with table formatting. The authors reviewed and revised all suggestions to ensure that they preserved the intended scientific content and retain full responsibility for the accuracy and integrity of the final manuscript.

\bibliography{custom}

% \clearpage

% \clearpage

\appendix

\section{Frequently Asked Questions (FAQs)}
\label{app:faqs}

\subsection{What are the main advantages of \sysname?}
\label{app:faq-advantages}

\sysname{} differs from conventional memory systems in how it maintains and uses personal information. Rather than storing facts independently, it organizes them within an ontology that tracks changes in validity, scope, and relations. This allows earlier records to remain accessible even when newer information supersedes or contradicts them.

The framework also separates memory storage from memory use. A learned policy generates a Reasoning State for each request, indicating which memories are relevant and how they should influence the response. The policy is trained using feedback from downstream answers, rather than relying on predefined rules or repeated self-reflection. The Reasoning State is provided as natural-language guidance and can be used with different response models without modifying their parameters.

\paragraph{A coherent and evolving memory state.}
The personal memory ontology organizes Persistent Facts and maintains their validity, scope, and relations as new interactions arrive. Original facts remain traceable, while relation-aware retrieval recovers the connected information needed for their interpretation.

\paragraph{Use-driven state evolution.}
For each request, a Reasoning State specifies the applicability, role, and priority of memory evidence. The Memory State Evolution Policy learns from whether candidate states improve the personal agent's answers, allowing experience from prior memory use to guide future requests.

\paragraph{Compatibility with different personal agents.}
Reasoning States are expressed in natural language and provided alongside the relevant facts and relations. They can therefore guide different personal-agent backbones without modifying their parameters.

\subsection{Why use a personal memory ontology instead of independent records?}
\label{app:faq-relation-completeness}

Independent memory records can preserve individual observations, but they provide limited information about how those observations relate over time. For example, a newly stated preference may replace an earlier one, while two conflicting records may need to remain unresolved until further evidence becomes available.

\sysname{} uses an ontology to represent these relationships explicitly. Facts are grouped under concepts and linked through typed relations, allowing retrieval to include not only a relevant record but also the information needed to interpret it. The original records are retained, so historical questions can still be answered after a preference or circumstance changes.

\paragraph{Connecting facts across interactions.}
The concept hierarchy organizes facts by topic, while their attributes and relations record changes in validity and scope. Earlier facts remain traceable even when they are superseded or conflict with later information.

\paragraph{Recovering complete evidence.}
Semantic retrieval may identify an individually relevant fact while omitting the related facts needed to interpret it. Starting from retrieved facts, \sysname{} traverses their stored relations to recover connected updates, conflicts, and constraints, providing the complete context required by the current request.

\subsection{Why are Reasoning States needed, and how do they differ from summaries?}
\label{app:faq-memory-use}

Relation States describe whether facts remain valid and how they are related, but do not determine how they should affect a particular request. A Reasoning State makes the applicability, role, and priority of memory evidence explicit for the current interaction.

\paragraph{Using the same fact in different contexts.}
A knee injury may answer a factual question, constrain an activity recommendation, or be irrelevant to another request. Similarly, a superseded preference may be irrelevant to a current decision but necessary for answering a historical question. The Reasoning State distinguishes these uses according to the request.

\paragraph{From memory content to response guidance.}
A Reasoning State is intended to guide the use of retrieved memories, rather than summarize their content. The distinction matters because the same fact can serve different purposes depending on the request.

Consider a user who previously mentioned a knee injury. The information is relevant when the user asks for exercise recommendations, but it may have no bearing on a question about travel expenses. Similarly, an outdated preference should not determine a current recommendation, although it remains useful when answering questions about the user’s past interests.

The Reasoning State captures these request-dependent distinctions. It is generated from the retrieved facts and their relations, while the original evidence remains available to the response model.

\begingroup
% Uses the manuscript's existing algorithm, algpseudocode, xcolor and amsmath packages.
\begin{algorithm*}[t]

\caption{Complete Workflow of \sysname{}}
\label{alg:statemem-inference}
\small
\renewcommand{\baselinestretch}{1.28}\selectfont
\algrenewcommand{\algorithmiccomment}[1]{\hfill\parbox[t]{5.4cm}{\raggedright\fontsize{7.8}{10}\selectfont\textcolor{black!65}{$\triangleright$~#1}\strut}}
\begin{algorithmic}[1]
\Require Dialogue history $\mathcal{H}_t$, ontology-backed training data $\mathcal{D}=\{(q_j,\mathcal{G}_j,y_j^*)\}$, Memory State Evolution Policy $\pi_\theta$, frozen personal agent $\mathcal{A}$, request $q$, seed count $K$, group size $G$
\Ensure Personal memory ontology $\mathcal{G}_t$, Persistent Facts $\mathcal{F}_t$, Relational States $\mathcal{R}_t$, trained policy $\pi_\theta$, answer $y$
\Statex \textbf{Phase 1: Ontology-Guided Memory State Induction}
\Statex \hfill\makebox[5.4cm][l]{\fontsize{7.8}{10}\selectfont\textcolor{black!65}{\textit{Explanation}}}
\State $\mathcal{F}_t=\{f_1,\ldots,f_{N_t}\} \gets \operatorname{ExtractTimeOrderedRecords}(\mathcal{H}_t)$ \Comment{Extract time-ordered Persistent Facts}
\State $\mathcal{G}^{(0)} \gets \operatorname{InitializeGraph}()$ \Comment{Initialize the personal memory ontology}
\For {$i=1,\ldots,N_t$} \Comment{Incorporate each Persistent Fact}
\State $\mathcal{G}^{(i)} \gets \operatorname{Induce}(\mathcal{G}^{(i-1)},f_i)$ \Comment{Update concepts, attributes, relations}
\EndFor \Comment{Finish ontology induction}
\State $\mathcal{G}_t \gets \mathcal{G}^{(N_t)}$ \Comment{Retain the induced ontology}
\State $\mathcal{R}_t \gets \{(f,a_f^{(N_t)},\rho_f^{(N_t)}) : f\in\mathcal{V}_F^{(N_t)}\}$ \Comment{Collect the Relational States}
\Statex
\Statex \textbf{Phase 2: RL-Based State Evolution}
\For {each policy optimization step} \Comment{Optimize the state evolution policy}
\State $(q_j,\mathcal{G}_j,y_j^*) \sim \mathcal{D}$ \Comment{Sample a training ontology and request}
\State $\pi_{\theta_{\mathrm{old}}} \gets \pi_\theta$ \Comment{Save the rollout policy snapshot}
\State $\mathcal{M}_j=(\mathcal{F}_j,\mathcal{R}_j) \gets \operatorname{AssembleEvidence}(q_j,\mathcal{G}_j,K)$ \Comment{Retrieve facts and Relational States}
\State $y_j^0 \gets \mathcal{A}(q_j,\mathcal{M}_j)$ \Comment{Answer without a Reasoning State}
\State $\{\mathcal{Z}_j^{(i)}\}_{i=1}^{G} \sim \pi_{\theta_{\mathrm{old}}}(\cdot\mid q_j,\mathcal{F}_j,\mathcal{R}_j)$ \Comment{Sample candidate Reasoning States}
\For {$i=1,\ldots,G$} \Comment{Evaluate each candidate Reasoning State}
\State $y_j^{(i)} \gets \mathcal{A}(q_j,\mathcal{M}_j,\mathcal{Z}_j^{(i)})$ \Comment{Answer using the Reasoning State}
\State $R_i \gets \operatorname{Score}(y_j^{(i)},y_j^*) - \operatorname{Score}(y_j^0,y_j^*)$ \Comment{Measure gain over the baseline}
\State $\widetilde{R}_i \gets \operatorname{ApplyGatesAndPenalties}(R_i,\mathcal{Z}_j^{(i)})$ \Comment{Check format and answer disclosure}
\EndFor \Comment{Finish candidate evaluation}
\State $\widehat{A}_{1:G} \gets \operatorname{NormalizeWithinGroup}(\widetilde{R}_{1:G})$ \Comment{Compute group-relative advantages}
\State $\pi_\theta \gets \operatorname{GRPOUpdate}\bigl(\pi_\theta,\pi_{\theta_{\mathrm{old}}},\{(\mathcal{Z}_j^{(i)},\widehat{A}_i)\}_{i=1}^{G}\bigr)$ \Comment{Update the policy with GRPO}
\EndFor \Comment{Finish policy optimization}
\Statex
\Statex \textbf{Phase 3: State-Aware Reasoning}
\State $(\mathcal{F}_q,\widetilde{\mathcal{R}}_q) \gets \operatorname{AssembleEvidence}(q,\mathcal{G}_t,K)$ \Comment{Retrieve facts and Relational States}
\State $\mathcal{Z}_q \gets \pi_\theta(q,\mathcal{F}_q,\widetilde{\mathcal{R}}_q)$ \Comment{Generate the Reasoning State}
\State $\mathcal{S}_t(q) \gets (\mathcal{R}_t,\mathcal{Z}_q)$ \Comment{Form the Memory State}
\State $y \gets \mathcal{A}(q,\mathcal{F}_q,\widetilde{\mathcal{R}}_q,\mathcal{Z}_q)$ \Comment{Answer using facts and both states}
\State \Return $\mathcal{G}_t,\mathcal{F}_t,\mathcal{R}_t,\pi_\theta,y$ \Comment{Return personal memory, policy, answer}
\end{algorithmic}
\end{algorithm*}
\endgroup

\subsection{Why use a separate policy model instead of agent self-reflection to generate Reasoning States?}
\label{app:faq-policy-model}

Self-reflection requires the personal agent to determine how memory should be used anew for each request, without retaining successful decisions from previous uses. In contrast, the Memory State Evolution Policy is trained offline by repeatedly generating and evaluating candidate Reasoning States, thereby internalizing memory-use patterns that improve downstream responses. At inference, it directly provides the personal agent with a ready-to-use Reasoning State, avoiding repeated execution or iterative reflection for each request. This separates the accumulation of memory-use experience from response generation and makes that experience reusable across requests and personal agents.

\subsection{How is the Memory State Evolution Policy trained and evaluated?}
\label{app:faq-evaluate-use}

The policy is trained according to how much its Reasoning States improve the personal agent's answers. Evaluation separately measures whether the system retrieves complete evidence and whether the personal agent uses that evidence correctly.

\paragraph{Learning from memory-use experience.}
Given a request, relevant Persistent Facts, and their Relation States, the policy samples a group of candidate Reasoning States. Each candidate is applied by a frozen personal agent under the same request and memory evidence, together with a baseline that uses the evidence directly. The resulting performance difference serves as feedback on the candidate's memory-use decisions and defines the reward in Equation~\ref{eq:relative-gain}. GRPO optimizes the policy using relative feedback within each group. Across repeated memory-use episodes, the policy internalizes reusable patterns for determining the applicability, role, and priority of memory evidence. Only the policy is updated, allowing this experience to guide the personal agent without modifying its parameters.

\paragraph{Separating evidence retrieval from evidence use.}
Complete Retrieval Rate measures how often the retrieved context contains all required evidence, while Acc@Complete measures answer accuracy on those cases. Figure~\ref{fig:memory_use} reports both metrics on LoCoMo-Refined and MemSyco-Bench, distinguishing errors caused by missing evidence from those caused by incorrect evidence use. Table~\ref{tab:component_ablation} further compares variants without Reasoning States or RL-based policy training. Together, these evaluations assess both the complete system and the contribution of learned state evolution.

\section{Method Details}
\label{app:method-details}

This section provides additional details for the three stages introduced in Section~\ref{sec:method}. We first present the complete formulation of Persistent Facts, Relational States, and Reasoning States. We then describe Ontology-Guided Memory State Induction, RL-Based State Evolution, and State-Aware Reasoning in the same order as the main text. Algorithm~\ref{alg:statemem-inference} summarizes memory construction, policy optimization, and inference.

\subsection{System Concepts}
\label{app:system-concepts}

Let $\mathcal{H}_t=\{h_1,\ldots,h_t\}$ denote the conversations accumulated through turn $t$, and let $q_t$ denote the current request.

\noindent\textbf{Definition 1 (Persistent Facts).}
The Persistent Facts
\begin{equation}
\small
\mathcal{F}_t=\{f_i\}_{i=1}^{N_t}
\end{equation}
preserve atomic information about the user's preferences, experiences, circumstances, beliefs, and constraints. Each fact retains its source, time, and evidence span. Later interactions do not overwrite these facts, but may update their associated states.

\noindent\textbf{Definition 2 (Relational States).}
Each fact $f_i$ is associated with a Relation State:
\begin{equation}
\small
r_i^t=(f_i,a_i^t,\rho_i^t),
\qquad
\mathcal{R}_t=\{r_i^t\mid f_i\in\mathcal{F}_t\},
\end{equation}
where $a_i^t$ specifies the current validity and scope of $f_i$, and $\rho_i^t$ records how it updates, conflicts with, or constrains other facts. The collection $\mathcal{R}_t$ forms the Relational States retained with personal memory. These states persist across requests and evolve when new conversations update the memory.

\noindent\textbf{Definition 3 (Reasoning State).}
For request $q_t$, the Reasoning State is inferred jointly from the request, Persistent Facts, and their Relational States:
\begin{equation}
\small
\mathcal{Z}_{q_t}
=
\Psi(q_t,\mathcal{F}_t,\mathcal{R}_t),
\end{equation}
where $\mathcal{Z}_{q_t}$ specifies the applicability, role, and priority of memory evidence. The state inference function $\Psi$ may be implemented by the personal agent or a separate model. In \sysname{}, it is instantiated by the Memory State Evolution Policy $\pi_\theta$.

\noindent\textbf{Definition 4 (Memory State).}
The Memory State for request $q_t$ combines the persistent Relational States with the request-specific Reasoning State:
\begin{equation}
\small
\mathcal{S}_t(q_t)
=
\left(
\mathcal{R}_t,
\mathcal{Z}_{q_t}
\right),
\qquad
\mathcal{M}_t(q_t)
=
\left(
\mathcal{F}_t,
\mathcal{S}_t(q_t)
\right).
\end{equation}
Persistent Facts specify what the memory contains, Relational States indicate whether these facts remain valid and how they are related, and the Reasoning State determines how they should be used for the current request.

\subsection{Ontology-Guided Memory State Induction}
\label{app:ontology-construction}

Memory construction proceeds incrementally as new dialogue sessions become available. For each session, the extractor receives the dialogue and a relevant branch of the existing ontology. It identifies atomic facts and proposes changes to the concept hierarchy and memory relations.

\begin{figure*}[t]
\centering
\includegraphics[width=\textwidth]{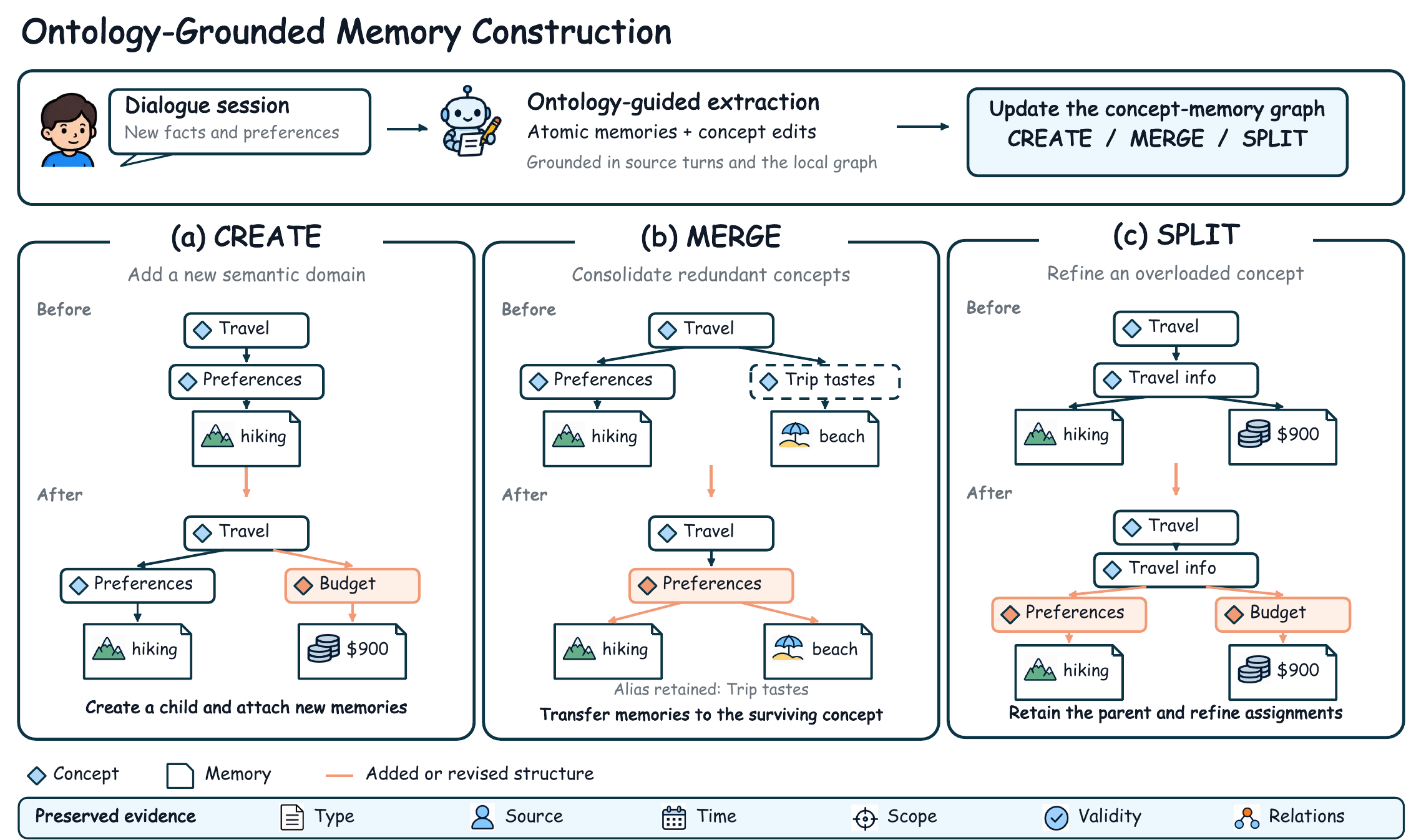}
\caption{Ontology-Guided Memory State Induction. Each dialogue session yields Persistent Facts, concept edits, and memory relations. \texttt{CREATE} adds a concept, \texttt{MERGE} combines duplicate concepts, and \texttt{SPLIT} introduces more specific concepts. Persistent Facts remain traceable while their Relational States evolve with the ontology.}
\label{fig:app-graph-build}
\end{figure*}

Each extracted fact retains its source turn, evidence span, timestamp, and qualifiers. Changes in validity or scope are recorded in its Relational State, leaving the original fact unchanged. Figure~\ref{fig:app-graph-build} illustrates the ontology operations used during this process.

\paragraph{Session-level fact extraction.}
For each new dialogue session, the extractor receives the session together with a relevant local ontology branch containing related concepts, neighboring hierarchy nodes, representative facts, and their incident relations. To keep the extraction prompt bounded as the ontology grows, each concept is embedded from its name, description, aliases, and recent active memories, and the top-5 closest concepts together with their root paths and active children are retrieved to form the local branch. It produces a time-ordered set of Persistent Facts and proposes the corresponding concept and relation updates. Each fact contains a single proposition and retains its record type, source turn, evidence span, explicit time, and qualifiers. Its current validity and scope are maintained in the associated Relation State rather than encoded by overwriting the fact. The four scope levels (\texttt{global}, \texttt{domain}, \texttt{task}, and \texttt{situation}) span broad to narrow applicability---from persistent user traits to one-off contexts---preventing local or situational habits from leaking globally.

\paragraph{Ontology construction and validation.}
Starting from an empty ontology $\mathcal{G}^{(0)}$, facts are incorporated in temporal order:
\begin{equation}
\small
\mathcal{G}^{(i)}
=
\operatorname{Induce}
\left(
\mathcal{G}^{(i-1)},
f_i
\right),
\end{equation}
where
\begin{equation}
\small
\mathcal{G}^{(i)}
=
\left(
\mathcal{V}_C^{(i)}
\cup
\mathcal{V}_F^{(i)},
\mathcal{E}_H^{(i)}
\cup
\mathcal{E}_R^{(i)}
\right).
\end{equation}
Here, $\mathcal{V}_C^{(i)}$ and $\mathcal{V}_F^{(i)}$ denote concept and fact nodes. Structural edges $\mathcal{E}_H^{(i)}$ use \texttt{PARENT\_OF} to form the concept hierarchy and \texttt{CONTAINS} to attach each fact to its most specific concept. The operations \texttt{CREATE}, \texttt{MERGE}, and \texttt{SPLIT} revise the hierarchy while preserving one parent for each nonroot concept and one concept assignment for each fact.

Validation ensures that extracted facts are grounded in the dialogue and that proposed edits preserve graph consistency. A \texttt{SUPERSEDES} edge connects a replacement fact to an earlier version and marks the earlier fact as superseded. A \texttt{CONFLICTS\_WITH} edge preserves unresolved inconsistencies and marks active endpoints as disputed. Relation endpoints must refer to compatible information types, and update relations must remain acyclic. Each session-level update is committed only after its facts, concepts, and relations pass validation.

\paragraph{Relational State induction.}
After incorporating the first $i$ facts, their Relational States are
\begin{equation}
\small
\mathcal{R}^{(i)}
=
\left\{
r_f^{(i)}
=
\left(
f,
a_f^{(i)},
\rho_f^{(i)}
\right)
\,\middle|\,
f\in\mathcal{V}_F^{(i)}
\right\},
\label{eq:ontology-relation-states}
\end{equation}
where $a_f^{(i)}$ records the current validity and scope of $f$, and $\rho_f^{(i)}$ contains its concept attachment and relations to other facts. New facts may revise these attributes and relations without overwriting the underlying fact. After all facts have been incorporated,
\begin{equation}
\small
\mathcal{R}_t=\mathcal{R}^{(N_t)}.
\end{equation}
The resulting Persistent Facts $\mathcal{F}_t$ and Relational States $\mathcal{R}_t$ form the persistent personal memory.

\subsection{RL-Based State Evolution}
\label{app:hint-interface}

Relational States describe how facts currently hold and relate, but do not specify how they should be used for a particular request. \sysname{} instantiates the state inference function $\Psi$ with a Memory State Evolution Policy $\pi_\theta$, which internalizes reusable memory-use experience through repeated generation and evaluation of candidate Reasoning States.

\paragraph{Reasoning State interface.}
For a request $q$, the policy receives the relevant Persistent Facts $\mathcal{F}_q$ and their Relational States $\mathcal{R}_q$:
\begin{equation}
\small
\mathcal{Z}_q
\sim
\pi_\theta
\left(
\cdot
\mid
q,
\mathcal{F}_q,
\mathcal{R}_q
\right).
\end{equation}
The input includes the facts, their validity and scope, temporal information, and relation annotations. The generated Reasoning State specifies which evidence applies, the role it should play, and how it should be prioritized. It is supplied alongside the original memory evidence so that the personal agent retains access to the facts supporting the guidance.

\paragraph{Learning from memory-use experience.}
For each training instance, the policy samples $G$ candidate Reasoning States:
\begin{equation}
\small
\mathcal{Z}_q^{(i)}
\sim
\pi_{\theta_{\mathrm{old}}}
\left(
\cdot
\mid
q,
\mathcal{F}_q,
\mathcal{R}_q
\right),
\qquad i=1,\ldots,G.
\end{equation}
Each candidate is applied by a frozen personal agent under the same request and memory evidence. A shared baseline uses the same evidence without a Reasoning State. Their performance difference provides feedback on the candidate's memory-use decisions and defines the reward in Equation~\ref{eq:relative-gain}. Format validation and penalties for revealing the reference answer further constrain the reward.

GRPO~\citep{shao2024deepseekmath} normalizes rewards within each sampled group and optimizes the clipped objective in Equation~\ref{eq:grpo}. Only $\pi_\theta$ is updated; memory construction, retrieval, scoring, and the personal agent remain fixed. For each training instance, the policy samples multiple candidate Reasoning States. Each candidate is evaluated using the same request and memory evidence, with the response model kept frozen. The reward is calculated as the difference between the answer score obtained with the candidate state and the score obtained without it.

Candidates are also checked for formatting violations and reference-answer disclosure. GRPO then updates the policy using rewards normalized within each group. The memory construction pipeline, retrieval procedure, and response model remain fixed throughout training.

The resulting policy is used to generate Reasoning States at inference time.

\subsection{State-Aware Reasoning}
\label{app:candidate-selection}

At inference, State-Aware Reasoning retrieves the Persistent Facts and Relational States required by the current request, generates a Reasoning State, and grounds the personal agent in both states.

\paragraph{Seed retrieval and relation traversal.}
Given request $q$, embedding similarity retrieves $K$ Persistent Facts as seeds. Because a seed may have been updated, contradicted, or constrained by other facts, relation traversal expands the initial set:
\begin{equation}
\small
\mathcal{F}_q
=
\bigcup_{
f\in
\operatorname{Retrieve}_{\mathrm{sim}}
(q,\mathcal{F}_t,K)
}
\operatorname{Traverse}
\left(
f,
\mathcal{R}_t
\right).
\end{equation}
Traversal follows stored memory relations in both directions until all connected facts have been collected. Earlier facts, replacements, and conflicting counterparts are retained with their original relation types and directions. Facts reached from multiple seeds are included only once, while isolated seeds remain individual facts.

\paragraph{Relational State assembly.}
The Relational States associated with the retrieved facts are
\begin{equation}
\small
\widetilde{\mathcal{R}}_q
=
\left\{
r_f^t\in\mathcal{R}_t
\mid
f\in\mathcal{F}_q
\right\}.
\end{equation}
Connected facts retain their validity, scope, provenance, temporal information, and relation annotations. No additional filtering is applied after traversal. The seed count $K$ controls the initial retrieval set, while the final context size depends on the relations connecting those seeds to other facts.

\paragraph{Reasoning and response generation.}
The learned policy generates a Reasoning State from the request, retrieved facts, and their Relational States. The personal agent then produces its response using the same memory evidence and the generated state:
\begin{equation}
\small
\begin{aligned}
\mathcal{Z}_q
&=
\pi_\theta
\left(
q,
\mathcal{F}_q,
\widetilde{\mathcal{R}}_q
\right),\\
y
&=
\mathcal{A}
\left(
q,
\mathcal{F}_q,
\widetilde{\mathcal{R}}_q,
\mathcal{Z}_q
\right).
\end{aligned}
\end{equation}
Relation traversal preserves the context needed to interpret the retrieved facts, while the Reasoning State specifies how the personal agent should use them for the current request.

\paragraph{Overall procedure.}
\label{app:inference-algorithm}
Algorithm~\ref{alg:statemem-inference} provides the full training and inference workflow. During inference, the system retrieves relevant facts, expands the evidence through stored relations, and generates a Reasoning State before passing the assembled context to the response model.

% Required packages:
% \usepackage{booktabs}
% \usepackage{graphicx}
% \usepackage[table]{xcolor}

% Required packages:
% \usepackage{booktabs}
% \usepackage{graphicx}
% \usepackage[table]{xcolor}

\subsection{Preliminary Case Analysis}
\label{app:preliminary-case}

\begin{figure}[t]
    \centering
    \includegraphics[width=\columnwidth]{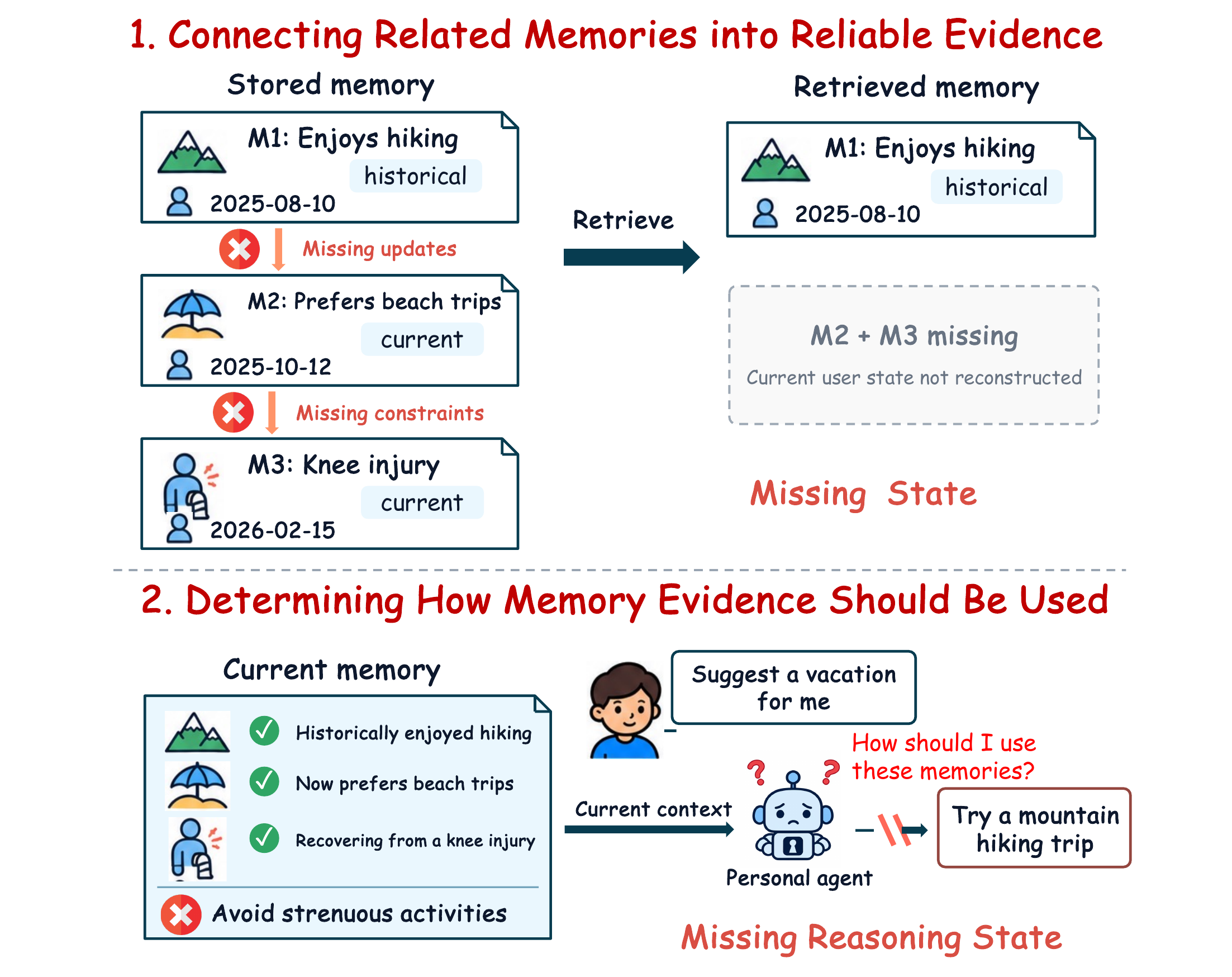}
    \caption{Case analysis of the two limitations identified in our preliminary studies. The upper panel shows that retrieving an individually relevant fact may omit related updates and constraints required to reconstruct the current memory state. The lower panel shows that even complete evidence may be used incorrectly when its applicability, role, and priority are not determined for the current request.}
\label{fig:preliminary-case}
\end{figure}

\textbf{Connecting related memories into reliable evidence.}
The upper panel of Figure~\ref{fig:preliminary-case} shows three memories accumulated over time. The user historically enjoyed hiking (\textit{M1}), later developed a preference for beach trips (\textit{M2}), and is currently recovering from a knee injury (\textit{M3}). A fact-centric system may retrieve only \textit{M1} because hiking is semantically relevant to a vacation request. However, \textit{M1} cannot be interpreted correctly in isolation: \textit{M2} updates the user's current preference, while \textit{M3} introduces a constraint against strenuous activities. Omitting either memory prevents the system from reconstructing the current memory state and may lead it to recommend an outdated and physically unsuitable hiking trip. This case illustrates that individually relevant facts do not necessarily form reliable evidence; their updates and constraints must be considered together.

\textbf{Determining how memory evidence should be used.}
The lower panel assumes that all three memories are available when the user asks for a vacation recommendation. Complete evidence, however, does not by itself determine the response. For this request, \textit{M1} should be treated as historical information, \textit{M2} as the current preference, and \textit{M3} as a constraint that excludes strenuous activities. A suitable response should therefore prioritize a low-intensity beach trip. Nevertheless, a personal agent may overemphasize the historically salient hiking preference and recommend a mountain hiking trip despite having access to the complete evidence. The same historical preference could be relevant to a question about the user's past activities, but should not guide the current recommendation. This is therefore a memory-use failure rather than a retrieval failure, showing that the applicability, role, and priority of each memory must be determined for the current request.

\begingroup

\def\TBHead#1{%
  \shortstack[c]{#1}%
}

\begin{table*}[t]
\centering

\fontsize{6.4}{7.0}\selectfont
\renewcommand{\arraystretch}{1.08}
\setlength{\tabcolsep}{1.3pt}

\resizebox{\linewidth}{!}{%
\begin{tabular}{@{}l*{12}{c}@{}}
\toprule

\textbf{Method}
& \multicolumn{5}{c}{\textbf{MemSyco-Bench}}
& \multicolumn{2}{c}{\textbf{LoCoMo-Refined}}
& \multicolumn{5}{c}{\textbf{MemTrapBench}}
\\

\cmidrule(lr){2-6}
\cmidrule(lr){7-8}
\cmidrule(lr){9-13}

&
\TBHead{Scope\\Acc. $\uparrow$}
&
\TBHead{Objective\\Acc. $\uparrow$}
&
\TBHead{Personalized\\Acc. $\uparrow$}
&
\TBHead{Preference\\Change Acc. $\uparrow$}
&
\TBHead{Preference\\Conflict Acc. $\uparrow$}
&
\TBHead{LLM-Judge\\Score $\uparrow$}
&
\TBHead{Token F1\\$\uparrow$}
&
\TBHead{Safety\\Acc. $\uparrow$}
&
\TBHead{Cognitive Bias\\Acc. $\uparrow$}
&
\TBHead{Task Boundary\\Acc. $\uparrow$}
&
\TBHead{Trauma\\Acc. $\uparrow$}
&
\TBHead{Avg.\\$\uparrow$}
\\

\specialrule{0.5pt}{2pt}{3pt}

\textbf{Mem0}
& \textbf{14.3}
& 34.7
& \underline{57.7}
& 54.5
& 22.3
& \textbf{25.7}
& \textbf{26.9}
& 56.8
& \underline{69.3}
& \textbf{81.9}
& 85.5
& 73.4
\\

\quad + Self-Recheck~\citep{hu2026op}
& 13.0
& \textbf{43.3}
& \textbf{60.5}
& \underline{64.7}
& \underline{25.0}
& 23.4
& 12.6
& \textbf{75.1}
& 55.5
& 76.4
& \textbf{94.6}
& \underline{75.4}
\\

\quad + MemGate~\citep{zhang2026beyond}
& \underline{13.7}
& 39.0
& 51.0
& 51.2
& 23.7
& \underline{24.2}
& 11.4
& 54.6
& 40.9
& \underline{78.7}
& 93.5
& 66.9
\\

\quad + Dynamic Partition~\citep{hannoon2026mitigating}
& 11.7
& 39.0
& 56.5
& 53.8
& 8.7
& 20.8
& 10.6
& \underline{69.2}
& 44.6
& 78.4
& 93.1
& 71.3
\\

\rowcolor{blue!8}
\textbf{\quad + Reasoning State (Ours)}
& 11.0
& \underline{39.3}
& 48.8
& \textbf{72.7}
& \textbf{42.0}
& 23.9
& \underline{24.9}
& 61.1
& \textbf{72.3}
& \textbf{81.9}
& \underline{93.8}
& \textbf{77.3}
\\

\bottomrule
\end{tabular}%
}
\caption{Comparison of memory use strategies with Mem0 across three benchmarks. Dynamic Rules (Ours) uses the learned Memory State Evolution Policy. \textbf{Bold} and \underline{underlined} values indicate the best and second best results, respectively.}
\label{tab:memsyco_five_categories}
\end{table*}
\endgroup

\section{Additional Experiments}
\label{app:additional-experiments}

We compare memory use strategies and guidance sources, examine representative responses, and analyze retrieval cost and quality. Additional experiments characterize the stored memory structure and the use of complete retrieved evidence. Together, these results examine how the Relational State supports memory access and the Reasoning State guides memory application.

\subsection{Comparison of Memory Use Strategies}
\label{app:memory-use-categories}

To compare memory application strategies, we augment Mem0, which extracts and consolidates personal facts, with Self-Recheck, MemGate, Dynamic Partition, or our generated Reasoning States. MemGate filters memory representations according to the query~\citep{zhang2026beyond}, while Dynamic Partition organizes memories into inferred categories at inference time~\citep{hannoon2026mitigating}. Table~\ref{tab:memsyco_five_categories} reports results on MemSyco-Bench, LoCoMo-Refined, and MemTrapBench, with our Reasoning States, expressed as dynamic memory rules, denoted by Dynamic Rules (Ours).

The learned rules are most effective when the response requires revising the influence of prior information. They achieve 72.7\% accuracy on preference changes and 42.0\% on preference conflicts, exceeding the strongest alternatives by 8.0\% and 17.0\%. On MemTrapBench, they obtain the highest average score of 77.3. Cognitive Bias accuracy reaches 72.3\%, exceeding Mem0's 69.3\% by 3.0\%, while Task Boundary accuracy matches Mem0 at 81.9\%. These scenarios require deciding whether an earlier preference, strategy, or task rule still applies. Dynamic rules express these decisions through temporal validity, scope, and priority, allowing the agent to use personal history while adapting its response to the current request.

MemGate and Dynamic Partition do not improve Task Boundary accuracy over Mem0, scoring 78.7\% and 78.4\%, respectively, compared with Mem0's 81.9\%. Their effects on preference updates and conflicts are also mixed. MemGate lowers preference change accuracy from 54.5\% to 51.2\%, while Dynamic Partition lowers preference conflict accuracy from 22.3\% to 8.7\%. These results suggest that filtering or grouping records alone does not reliably determine which information should govern the response. Our rules address this decision explicitly, with training feedback based on answer improvement. Self-Recheck leads on objective judgment, personalized recommendations, Safety, and Trauma, while Mem0 retains the best scope and LoCoMo-Refined results. The clearest advantage of learned rules lies in handling changing preferences, conflicting information, and cognitive bias.

\subsection{Guidance from the Answer Model}
\label{app:self-gate-comparison}

\begin{figure*}[t]
    \centering
    \includegraphics[width=\textwidth]{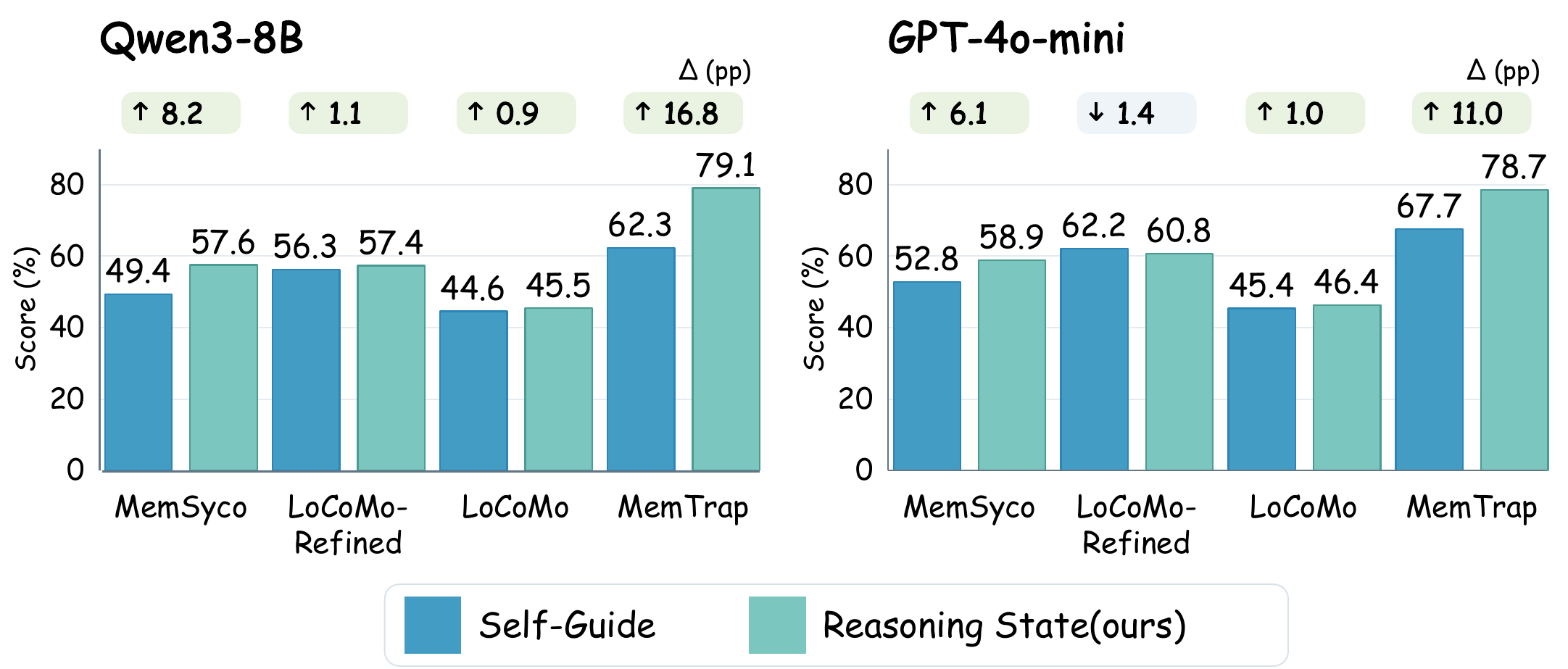}
    \caption{Performance with response model guidance (Self-Guide) and learned memory rules (Reasoning State) under Qwen3-8B and GPT-4o-mini. LoCoMo and LoCoMo-Refined report LLM judge scores. Differences from Self-Guide are reported in points on the percentage scale.}
    \label{fig:self_gate_comparison}
\end{figure*}

To compare learned memory-use guidance with guidance generated by the response model, we evaluate Self-Guide and Dynamic Rules under Qwen3-8B and GPT-4o-mini (Figure~\ref{fig:self_gate_comparison}). Self-Guide uses the response model to generate guidance, whereas Dynamic Rules uses the trained Memory State Evolution Policy to generate a Reasoning State.

Dynamic Rules yields larger gains on memory-use benchmarks than on conversational recall. Relative to Self-Guide, it improves MemSyco-Bench accuracy by 8.2 percentage points with Qwen3-8B and 6.1 percentage points with GPT-4o-mini, with corresponding MemTrapBench score gains of 16.8 and 11.0 points. Improvements on LoCoMo are smaller, at 0.9 and 1.0 points, respectively. On LoCoMo-Refined, Dynamic Rules improves the Qwen3-8B score by 1.1 points but reduces the GPT-4o-mini score by 1.4 points. These results suggest that learned guidance is particularly useful for determining the applicability and priority of memories, while offering less consistent benefits for conversational recall.

Conversational recall primarily requires recovering and combining stored information. MemSyco-Bench and MemTrapBench also test whether historical information should influence the current response, particularly when preferences change or prior interactions introduce misleading assumptions. Unlike Self-Guide, the Memory State Evolution Policy is trained using feedback on the answers produced with its guidance. Candidate rules are evaluated against a baseline answer generated from the same records without rules, assessing whether applying, restricting, or prioritizing memories improves correctness. The larger gains on memory-use benchmarks support learning from memory-use experience to determine the applicability, role, and priority of evidence.

\subsection{Case Study}
\label{app:case-study}

\begin{figure*}[t]
    \centering
    \includegraphics[width=\textwidth]{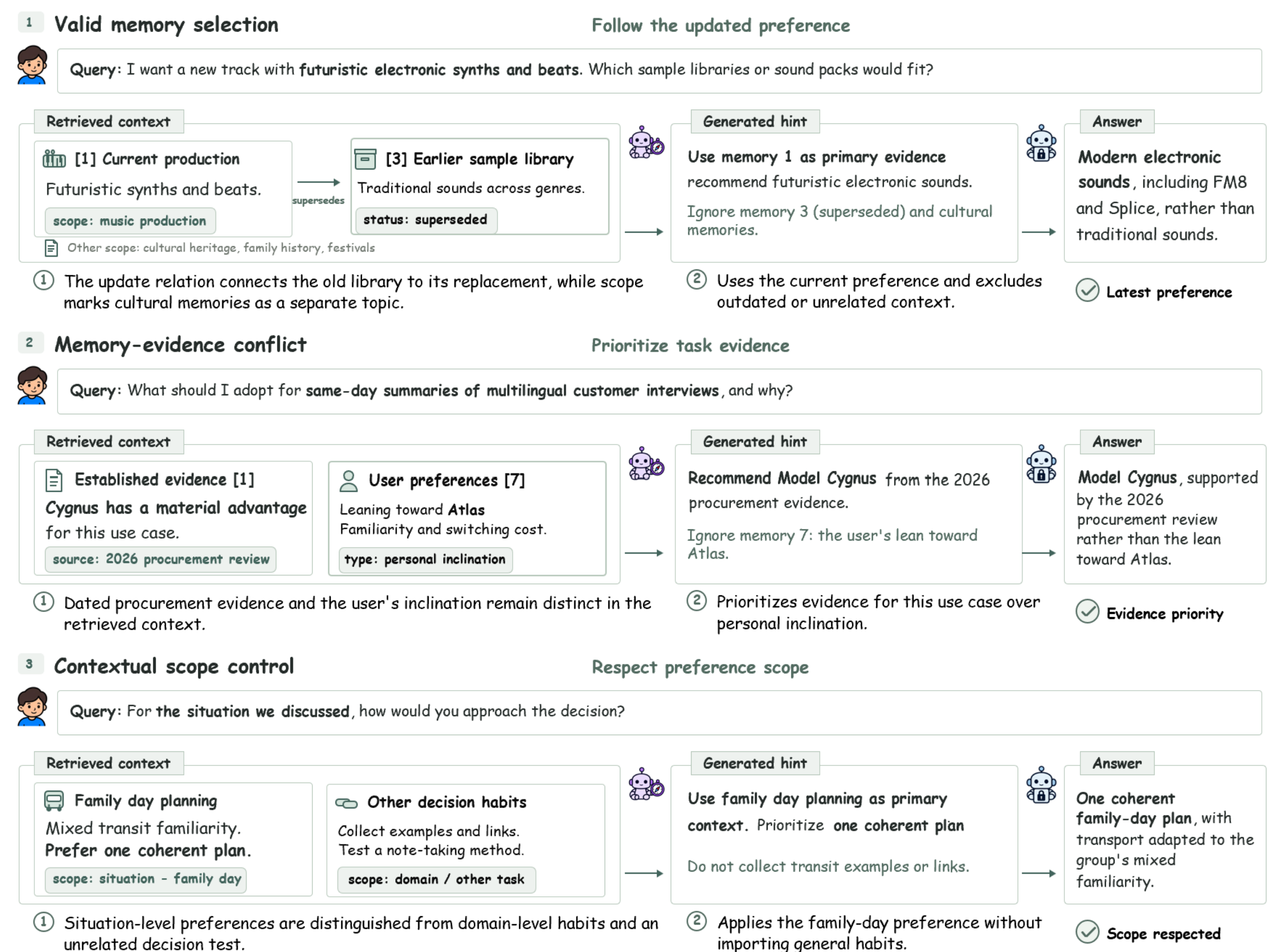}
    \caption{Examples of Reasoning States expressed as dynamic memory rules for preference updates, evidence priority, and scope control. Each case presents the relevant evidence, generated rules, and response. Text is condensed from recorded outputs.}
    \label{fig:app-case-study}
\end{figure*}

Figure~\ref{fig:app-case-study} presents three examples from MemSyco-Bench, covering preference updates, evidence priority, and scope control. Each example includes the retrieved memories, the generated Reasoning State, and the final response. We examine whether the generated rules correctly distinguish outdated preferences from current ones, prioritize task-relevant evidence, and apply user information within its intended scope.

\paragraph{Following an updated preference.}
The user requests futuristic electronic sounds for a new track. The retrieved records connect this requirement to the earlier sample library through an update relation and distinguish music production from cultural interests through scope. The generated rule prioritizes the current production requirement and excludes the superseded library and unrelated cultural memories. The response recommends electronic sounds and tools, including FM8 and Splice. The example shows how an explicit update can change personalization for the current task while the earlier preference remains stored.

\paragraph{Prioritizing evidence for the requested task.}
The user asks which system to adopt for same day summaries of multilingual customer interviews. The records distinguish the user's inclination toward Atlas from a 2026 procurement assessment favoring Cygnus. The rule assigns priority to the procurement evidence, and the response recommends Cygnus. Here, the source of a record matters to its role in the decision. Preserving the distinction between personal inclination and task evidence allows the rule to specify their relative priority.

\paragraph{Applying preferences within their scope.}
The request refers to a previously discussed situation. The records identify family day planning and distinguish the preference for one coherent plan from broader habits of collecting examples and links. The rule applies the planning preference to the current decision, and the response organizes the activities into one plan with transport adapted to the group's differing transit familiarity. Scope connects the remembered preference to the situation in which it should influence the answer.

\subsection{Efficiency Analysis}
\label{app:efficiency-analysis}

\begin{figure}[t]
    \centering
    \includegraphics[width=0.5\textwidth]{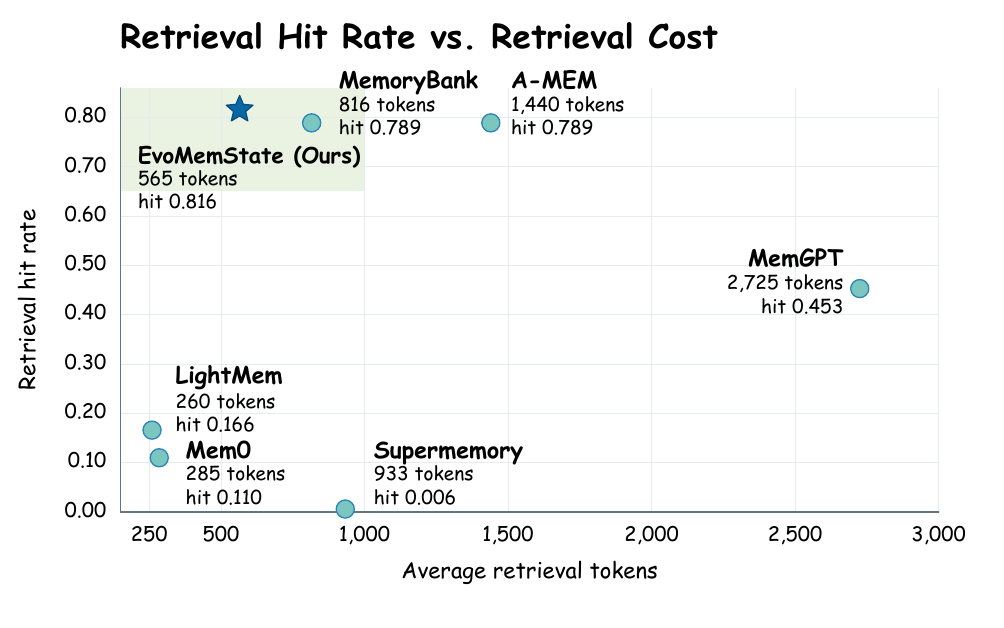}
    \caption{Retrieval hit rate and average retrieved context length on LoCoMo-Refined. Each point represents a memory system, and retrieval cost is measured in tokens.}
    \label{fig:hit_cost_rate}
\end{figure}

To assess retrieval efficiency, Figure~\ref{fig:hit_cost_rate} relates evidence access to the average number of retrieved context tokens on LoCoMo-Refined. For $N$ questions, hit rate is $\frac{1}{N}\sum_{i=1}^{N}\mathbb{I}[n_i^{\mathrm{hit}}\geq 1]$, where $n_i^{\mathrm{hit}}$ counts retrieved gold evidence items. A question counts as a hit when at least one gold item is retrieved, irrespective of how much other required evidence is missing. Context length measures the tokens supplied to the response model.

\sysname{} achieves the highest hit rate of 0.816 with an average context of 565 tokens. MemoryBank retains dialogue history and layered summaries, while A-MEM represents interactions as linked contextual notes. Both reach 0.789 with 816 and 1,440 tokens, respectively. Our context is 30.8\% shorter than MemoryBank's and 60.8\% shorter than A-MEM's, while recovering gold evidence for a larger proportion of questions.

One possible explanation is the difference in memory representation. MemoryBank and A-MEM retain more contextual information within individual records, whereas \sysname{} stores facts separately and connects them through explicit relations. This allows the system to retrieve related information without necessarily including the surrounding dialogue or longer memory notes.

LightMem, which compresses dialogue segments into summaries, and Mem0, which stores compact extracted memories, supply shorter contexts but attain hit rates of only 0.166 and 0.110. Their results illustrate the tension between reducing context length and maintaining access to evidence. The advantage of \sysname{} lies in its combination of compact records and access to associated information, yielding more frequent evidence hits at a lower context cost than the strongest retrieval baselines. Hit rate establishes whether retrieval finds an initial piece of evidence, while coverage and complete retrieval rate in Table~\ref{tab:retrieval_quality} examine how much of the required set is recovered.

\begin{table}[t]
\centering
\scriptsize
\setlength{\tabcolsep}{3.0pt}
\renewcommand{\arraystretch}{1.12}
\resizebox{\columnwidth}{!}{%
\begin{tabular}{lrrrrr}
\toprule
\textbf{System} & \textbf{Input} & \textbf{Output} & \textbf{Total} & \textbf{Cost (\$)} & \textbf{Time} \\
\midrule
MemoryBank & 33.8K & 1.7K & 35.5K & 0.011 & 169s \\
SimpleMem & 29.7K & 51.5K & 81.2K & 0.039 & 2,622s \\
LightMem & 287.0K & 42.7K & 329.7K & 0.104 & 9,503s \\
Mem0 & 774.6K & 110.9K & 885.5K & 0.288 & 8,186s \\
A-MEM & 612.8K & 199.9K & 812.7K & 0.296 & 13,834s \\
\midrule
\rowcolor{blue!8}
\textbf{\sysname{} (Ours)} & \textbf{211.1K} & \textbf{49.4K} & \textbf{260.5K} & \textbf{0.091} & \textbf{632s} \\
\bottomrule
\end{tabular}%
}
\caption{Offline memory construction overhead on LoCoMo \texttt{conv-26} (19 sessions, 419 turns) using \texttt{deepseek-v4-flash}.}
\label{tab:build_efficiency}
\end{table}

Beyond retrieval context length, Table~\ref{tab:build_efficiency} compares offline memory construction overhead on LoCoMo \texttt{conv-26} (19 sessions, 419 turns) using \texttt{deepseek-v4-flash}. Session-level induction over a bounded local ontology branch avoids turn-by-turn rewriting, requiring 21\%--71\% fewer total tokens, 1.1$\times$--3.3$\times$ lower API cost, and 13$\times$--22$\times$ shorter wall-clock build time than LightMem, Mem0, and A-MEM (4.2$\times$ faster than SimpleMem), while per-session input prompts stay flat at $\sim$12.5K tokens in sessions 10--19. Although MemoryBank costs less to build, it trails \sysname{} by 10.0 points overall with Qwen3-8B in Table~\ref{tab:main-results}. At inference time, the 4B policy adds 94.7 generated tokens per query ($<$15\% runtime overhead), and sparse relation traversal ($K=10$) takes 3.08\,$\mu$s on MemSyco (22 nodes), 5.97\,$\mu$s on LoCoMo (621 nodes), and 3.09\,$\mu$s on synthetic graphs with 50K nodes (0.03 edges/node).

\subsection{Retrieval Quality}
\label{app:retrieval-quality}

To distinguish partial evidence recovery from complete retrieval, Table~\ref{tab:retrieval_quality} reports three metrics on LoCoMo-Refined. For question $i$, $n_i^{\mathrm{gold}}$ denotes the number of gold evidence items and $n_i^{\mathrm{hit}}$ the number retrieved. Hit rate, coverage, and complete retrieval rate average $\mathbb{I}[n_i^{\mathrm{hit}}\geq 1]$, $n_i^{\mathrm{hit}}/n_i^{\mathrm{gold}}$, and $\mathbb{I}[n_i^{\mathrm{hit}}=n_i^{\mathrm{gold}}]$, respectively, across questions. Hit rate counts questions with any gold evidence, coverage measures the fraction recovered per question, and complete retrieval rate requires the entire gold set.

\sysname{} leads on all three metrics, with a hit rate of 0.816, coverage of 0.747, and complete retrieval rate of 0.680. MemoryBank, which retains history through layered summaries, and A-MEM, which stores linked contextual notes, achieve complete retrieval rate of 0.669 and 0.662. Their advantage over NaiveRAG's 0.564 is consistent with the importance of preserving context beyond individual dialogue fragments. The ontology represents this context through atomic records, metadata, and explicit relations. Retrieval can therefore recover associated updates and conflicts even when those records are not among the initial similarity matches, supporting interpretation of the retrieved information.

Across systems, complete retrieval rate remains below hit rate. For \sysname{}, the difference is 13.6 percentage points, corresponding to questions where retrieval finds part of the gold evidence but omits the rest. An answer that depends on several dialogue turns may require details that are less similar to the query than the first retrieved record. Coverage quantifies these partial recoveries, while complete retrieval rate tests whether every required item is available. Retrieving a relevant fact alone can therefore leave out the updates, conflicts, or constraints needed to interpret it.

\begin{table}[t]
\centering
\small
\setlength{\tabcolsep}{6.0pt}
\renewcommand{\arraystretch}{1.12}
\resizebox{\columnwidth}{!}{%
\begin{tabular}{lccc}
\toprule
\textbf{Method}
& \shortstack{\textbf{Retrieval}\\\textbf{Hit Rate}}
& \shortstack{\textbf{Average}\\\textbf{Coverage}}
& \shortstack{\textbf{Complete}\\\textbf{Retrieval Rate}} \\
\midrule
\rowcolor{blue!8}
\textbf{\sysname{} (Ours)}
& \textbf{0.816}
& \textbf{0.747}
& \textbf{0.680} \\
MemoryBank
& \underline{0.789}
& \underline{0.727}
& \underline{0.669} \\
A-MEM
& \underline{0.789}
& 0.724
& 0.662 \\
NaiveRAG
& 0.684
& 0.615
& 0.564 \\
MemGPT
& 0.453
& 0.384
& 0.323 \\
LightMem
& 0.166
& 0.144
& 0.129 \\
Mem0
& 0.110
& 0.095
& 0.086 \\
Supermemory
& 0.006
& 0.004
& 0.003 \\
\bottomrule
\end{tabular}%
}
\caption{Gold evidence recovery on LoCoMo-Refined. Hit rate counts questions with at least one gold item retrieved, coverage averages the fraction of gold items retrieved per question, and complete retrieval rate counts questions with every gold item retrieved. All three metrics are averaged across questions and reported on a 0--1 scale.}
\label{tab:retrieval_quality}
\end{table}

\begin{figure*}[t]
    \centering
    \includegraphics[width=\textwidth]{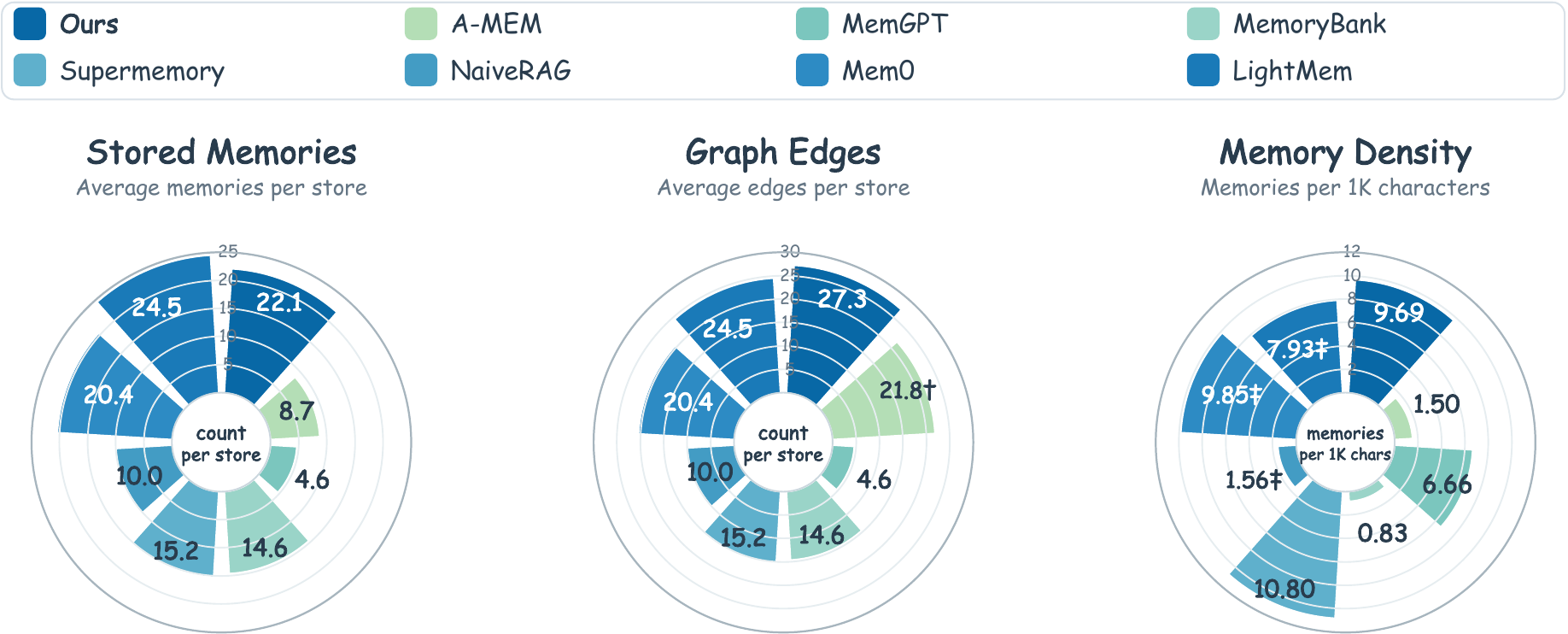}
    \caption{Statistics of memory stores on MemSyco-Bench. The panels report average entry counts, edge counts, and entries per thousand characters. Flat stores use entry counts as an edge count proxy. $\dagger$ includes cross note links in A-MEM. $\ddagger$ denotes density estimated from retrieved entries and context length.}
    \label{fig:memory-structure}
\end{figure*}

\subsection{Memory Structure Statistics}
\label{app:memory-structure}

To characterize the granularity and organization of personal memory, Figure~\ref{fig:memory-structure} reports statistics on MemSyco-Bench. The three panels show average record counts, average edge counts, and memory density, measured as records per thousand characters. Edge counts include concept and memory relations in our ontology and links among notes in A-MEM. Flat stores use record counts as an edge count proxy. Higher density indicates shorter records. Density is estimated from retrieved records and context length for NaiveRAG, Mem0, and LightMem, and computed directly from stored records for the remaining methods.

\sysname{} maintains an average of 22.1 records and 27.3 edges per store, with a record count comparable to Mem0's 20.4 and LightMem's 24.5. The ontology adds semantic organization and temporal relations to these atomic records. A-MEM connects contextual notes through untyped links, while our relation types encode concept membership, hierarchy, and memory updates. These semantics support distinct operations on the same personal history, placing new records within the profile and linking revised information to its earlier versions. Preserving both versions makes changes in the user's preferences explicit and leaves historical information available for later requests.

At 9.69 records per thousand characters, \sysname{} has a density close to the compact fact representations of Mem0 and Supermemory. Each record also retains an evidence quote, linking its concise statement to the source dialogue. A-MEM and MemoryBank store longer contextual notes and layered summaries, with densities of 1.50 and 0.83, respectively. Their context is carried largely within textual entries. Our ontology distributes this context across atomic records, source attributes, and explicit relations, allowing individual facts to remain concise while preserving how they should be interpreted together. The reported density therefore describes the granularity of the stored records rather than the total amount of information maintained by the system.

\subsection{Memory Application with Complete Evidence}
\label{app:complete-evidence-guidance}

\begin{figure*}[t]
    \centering
    \includegraphics[width=0.9\textwidth]{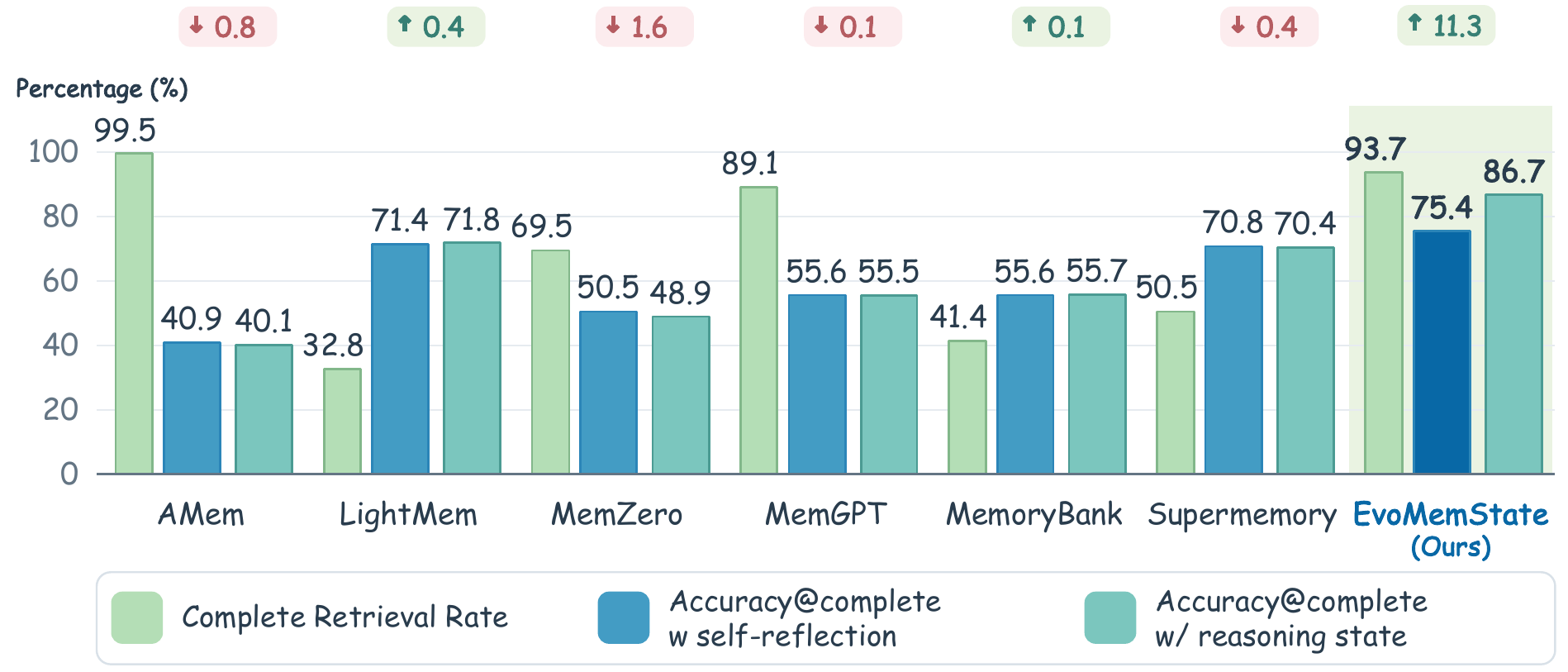}
    \caption{Complete retrieval rate and Acc@complete with and without memory guidance. Acc@complete is evaluated on questions with all required evidence retrieved. Arrows report changes in percentage points.}
    \label{fig:memory_use_app}
\end{figure*}

To assess how Reasoning States support memory application when retrieval supplies all required evidence, we evaluate the Valid Memory Selection tasks of MemSyco-Bench. Figure~\ref{fig:memory_use_app} reports complete retrieval rate and compares Acc@complete under two conditions: \textit{w/ self-reflection}, where the personal agent reflects on how to use the retrieved memories, and \textit{w/ reasoning state}, where it uses an explicit Reasoning State to guide memory application. Complete retrieval rate gives the proportion of questions with all gold evidence retrieved, while Acc@complete measures answer accuracy on this subset for each system.

\sysname{} achieves a complete retrieval rate of 93.7\%. Its Acc@complete increases from 75.4\% with self-reflection to 86.7\% with a Reasoning State, a gain of 11.3 percentage points. This improvement on questions with complete evidence suggests that retrieving the required records does not by itself ensure correct memory use. The agent must still resolve differences in temporal validity, scope, and evidence roles. The Reasoning State makes these application decisions explicit through dynamic rules, specifying how updates, constraints, and competing information affect the current request.

For the baselines, replacing self-reflection with a Reasoning State yields only small gains or decreases: +0.4 percentage points for LightMem, +0.1 for MemoryBank, -0.8 for A-MEM, -1.6 for Mem0, -0.1 for MemGPT, and -0.4 for Supermemory. Thus, an explicit Reasoning State alone does not consistently improve memory application. The larger gain for \sysname{} is consistent with the benefit of its Memory State Evolution Policy, which learns application decisions through downstream answer-quality feedback. These results suggest that effective memory use depends on the quality of the decisions encoded in the state, rather than merely adding an intermediate reflection or instruction step.

\subsection{Cross-System Transfer Analysis}
\label{app:transfer-analysis}

To examine why cross-system transfer gains in Figure~\ref{fig:hint_bridge_comparison} vary across baselines, Table~\ref{tab:transfer_analysis} relates each memory system's average complete retrieval rate (CRR) and retrieved context length to the performance gain from adding policy-generated Reasoning States.

\begin{table}[t]
\centering
\scriptsize
\setlength{\tabcolsep}{5.0pt}
\renewcommand{\arraystretch}{1.12}
\resizebox{\columnwidth}{!}{%
\begin{tabular}{lrrr}
\toprule
\textbf{System} & \textbf{CRR (\%)} & \textbf{Context (tok)} & \textbf{Gain ($\Delta$)} \\
\midrule
NaiveRAG & 75.4 & 316 & \textbf{+4.9} \\
MemoryBank & \textbf{86.4} & 816 & \textbf{+4.8} \\
A-MEM & 79.6 & 1,440 & +3.4 \\
Mem0 & 51.5 & 285 & +2.2 \\
MemGPT & 66.9 & 2,725 & +0.7 \\
LightMem & 40.6 & 260 & +0.2 \\
\bottomrule
\end{tabular}%
}
\caption{Complete retrieval rate (CRR), average retrieved context length on LoCoMo-Refined, and overall cross-system transfer gain ($\Delta$) from Figure~\ref{fig:hint_bridge_comparison} with Qwen3-8B.}
\label{tab:transfer_analysis}
\end{table}

As shown in Table~\ref{tab:transfer_analysis}, transfer gains depend jointly on retrieval completeness, context structure, and task type. NaiveRAG (75.4\% CRR, 316 tokens), MemoryBank (86.4\% CRR, 816 tokens), and A-MEM (79.6\% CRR, 1,440 tokens) preserve high evidence coverage in raw dialogue or contextual notes, leaving clear room for the Reasoning State to resolve memory roles and priority (+3.4 to +4.9 points). By contrast, LightMem loses key facts during aggressive compression (40.6\% CRR, +0.2), whereas MemGPT's long prompts (2,725 tokens) dilute routing instructions on short-answer queries (+0.7 overall despite +5.7 on MemTrapBench). Across memory systems, Reasoning States provide the largest benefits when retrieved evidence is complete yet evolving or conflicting, such as preference updates (+28.0 points), memory--evidence conflicts (+22.5 points), and cognitive traps.

\section{Implementation Details}
\label{app:implementation-details}

This section describes the training data and evaluation setup for \sysname{}. We first explain how data instances are constructed to train the Memory State Evolution Policy, then introduce the benchmark datasets, baselines, and evaluation metrics used to assess memory recall and use in personal agents.

\subsection{Training Data Construction}
\label{app:training-data-construction}
To train the Memory State Evolution Policy to generate a Reasoning State from a request and its relevant Relational State, we construct approximately 60K question–memory training instances from FEVER~\citep{thorne2018fever}, SciFact~\citep{wadden2020scifact}, and PersonaMem~\citep{jiang2025personamem}. For objective questions, we convert the fact-verification tasks in FEVER and SciFact into question-answering instances with supporting evidence and reference answers. We then construct user memories that conflict with the objective evidence from the perspectives of personal belief, identity, and perceived expertise. For preference questions, we use the preference annotations in PersonaMem to construct scenarios involving current preferences, historical preferences, and preference updates, enabling the model to learn how different memories should influence a response.

To improve the difficulty and diversity of the training data, we further introduce distractor memories. For objective questions, bias-related memories are mixed with irrelevant everyday information. For preference questions, we add one or two genuine preferences from unrelated topics associated with the same user, including more challenging distractors drawn from the same conversational scenario but concerning a different topic. This design requires the model to identify relevant information based on memory content and applicability, rather than relying on superficial cues such as memory count, position, or fixed rule templates.

\begin{table*}[!t]
\centering
\begingroup
\setlength{\smallskipamount}{2pt}
% Preamble: \usepackage{tcolorbox} and \tcbuselibrary{skins}.
\definecolor{promptGreen}{RGB}{120,195,120}
\tcbset{enhanced, drop shadow,
    colback=white, colframe=promptGreen, colbacktitle=promptGreen,
    coltext=black, coltitle=white, width=\textwidth,
    fontupper=\footnotesize,
    fonttitle=\Large\bfseries,
    sharp corners, boxrule=1.5mm, arc=0mm, outer arc=0mm, boxsep=1pt,
    left=3mm, right=3mm, top=2mm, bottom=2mm,
    toptitle=2mm, bottomtitle=2mm,
    before skip=0pt, after skip=5pt}
\begin{tcolorbox}[title=Memory Construction]
\textbf{System}\par
You construct one user's self-evolving ontological memory.
The input contains a new dialogue and a small branch retrieved from the user's existing ontology.
In ONE response, extract atomic memories and propose the minimal ontology evolution required to organize them.

\smallskip
This memory stores two different things, and both matter:\par
- what is true of the USER (preferences, beliefs, constraints, experiences)\par
- what is true of the WORLD, as established by evidence surfaced in the dialogue (facts)\par
Keep the second kind independent of the first. Evidence does not become less true because the user prefers otherwise, and it must survive in the memory on its own terms.

\smallskip
Fixed meta-ontology:\par
- memory type: preference | fact | belief | experience | constraint\par
- scope level: global | domain | task | situation\par
- memory status is managed by the system; every new memory starts active\par
- memory relation: SUPERSEDES | CONFLICTS\_WITH\par
- concept edits are limited to CREATE | MERGE | SPLIT

\smallskip
Return exactly one JSON object with this shape:\par
\{\par
\hspace*{0.5em}"concept\_edits": [ CREATE | MERGE | SPLIT ],\par
\hspace*{0.5em}"memories": [ atomic memory with type, scope, time, source, parent\_ref, confidence ],\par
\hspace*{0.5em}"relations": [ \{ "type": "SUPERSEDES", "source\_ref", "target\_ref", "reason" \} ]\par
\}

\smallskip
Rules:
\begin{enumerate}
\setlength{\itemsep}{0pt}\setlength{\topsep}{1pt}\setlength{\parsep}{0pt}\setlength{\parskip}{0pt}
\item Extract anything that could answer a later factual or temporal question about either speaker, not only durable preferences. This includes one-off events, dates, quantities, named entities, past actions, and biographical facts. Skip only pure greetings, backchannel, and meta-chatter with no propositional content.
\item A leaf is atomic: do not combine independent propositions. Preserve qualifiers, negation, scope, and time.
\item Reuse the narrowest suitable existing concept. CREATE only when no supplied concept fits.
\item CREATE a shallow, reusable concept, not a concept that paraphrases one memory.
\item MERGE only clear duplicate concepts. SPLIT only a genuinely overloaded concept.
\item Use SUPERSEDES when a newer state replaces an older state of the same kind; use CONFLICTS\_WITH for an unresolved contradiction. Do not relation-link mere topical similarity.
\item A SUPERSEDES pair must share the same memory type, but it need NOT share the same concept, scope, or wording.
\item When a newer preference replaces an older one, emit BOTH memories plus a SUPERSEDES from the newer to the older; never delete or rewrite the older memory.
\item Scope controls applicability. Use global only when the statement truly applies broadly.
\item Cite an exact evidence quote copied verbatim from a provided turn. Never cite a speaker's own unquoted prose from an assistant turn.
\item Quoted material is EXTERNAL EVIDENCE about the world. Record it as type fact, even when it contradicts the user.
\item Never let the user's stated preference decide which evidence is worth keeping.
\item Extract from EVERY speaker, not just one.
\item Preserve temporal anchors verbatim, and fill the time field.
\item Return up to 32 high-value atomic memories.
\item If there is genuinely no propositional content, return empty arrays. Output JSON only.
\end{enumerate}
\smallskip
\textbf{User input (fields)}\par
\texttt{\{new\_dialogue\}}\par
\texttt{\{local\_ontology\_branch\}}
\end{tcolorbox}
\endgroup
\caption{Prompt for extracting memories and updating the ontology.}
\label{tab:prompt-set}
\end{table*}

\subsection{Benchmark Datasets}
\label{app:benchmark-datasets}

We evaluate \sysname{} on four complementary benchmarks covering personal memory recall and use. LoCoMo and LoCoMo-Refined assess recall and reasoning about personal information across dialogue sessions. MemSyco-Bench tests whether agents appropriately apply user preferences, respect their scope, and track updates. MemTrapBench examines whether past interactions bias reasoning on the current task.

\paragraph{MemSyco-Bench.}
MemSyco-Bench~\citep{xiang2026memsyco} evaluates when and how agents should use user-specific memories in current responses. Personalized Memory Use and Valid Memory Selection test whether agents tailor recommendations to valid preferences and follow preference updates. Objective Fact Judgment, Contextual Scope Control, and Memory-Evidence Conflict test whether historical user beliefs or preferences improperly override facts, exceed their scope, or outweigh current evidence. Each instance introduces user information through a multi-turn dialogue, then tests its appropriate use in the final request. Together, these tasks assess personalization and memory-induced sycophancy in personal agents.

\paragraph{LoCoMo.}
LoCoMo~\citep{maharana2024locomo} evaluates recall and reasoning about speakers' experiences, relationships, and life events across dialogue sessions. Conversations are generated by agents grounded in speaker personas and temporal event graphs, then verified and edited by human annotators for consistency. We use its question-answering task, which covers single-hop retrieval, multi-hop reasoning, temporal reasoning, open-domain knowledge, and adversarial questions. These categories test whether agents recover personal details, connect information across sessions, and recognize queries unsupported by the history.

\paragraph{LoCoMo-Refined.}
LoCoMo-Refined\footnote{\url{https://github.com/mem-eval-suite/LoCoMo_refined}} recalibrates LoCoMo through corrected question-answer annotations and stricter answer evaluation. AI-assisted screening and human review revise ambiguous questions, reversed subject-object relations, and temporal inconsistencies with the source dialogues. The refined judge requires answers to cover the reference information, preserve temporal granularity, and avoid contradictions or unsupported additions. It provides a stricter assessment of factual and temporal accuracy when agents recall personal information from dialogue history.

\paragraph{MemTrapBench.}
MemTrapBench~\citep{fang2026memtrap} evaluates memory-induced cognitive traps through Reasoning Fixation and Belief Distortion. Reasoning Fixation comprises Task Boundary, Cognitive Bias, and Trauma, testing the persistence of prior task rules, overgeneralization of successful strategies, and avoidance induced by negative feedback. Belief Distortion uses Safety scenarios to test whether historical counterfactual premises override current safety judgments. Each final query is independently answerable, allowing comparison with generation without memory to measure the effect of historical context on current reasoning.

\begin{table*}[!t]
\centering
\begingroup
\setlength{\smallskipamount}{2pt}
% Preamble: \usepackage{tcolorbox} and \tcbuselibrary{skins}.
\definecolor{promptBlue}{RGB}{110,120,190}
\tcbset{enhanced, drop shadow,
    colback=white, colframe=promptBlue, colbacktitle=promptBlue,
    coltext=black, coltitle=white, width=\textwidth,
    fontupper=\footnotesize,
    fonttitle=\Large\bfseries,
    sharp corners, boxrule=1.5mm, arc=0mm, outer arc=0mm, boxsep=1pt,
    left=3mm, right=3mm, top=2mm, bottom=2mm,
    toptitle=2mm, bottomtitle=2mm,
    before skip=0pt, after skip=5pt}
\begin{tcolorbox}[title=Reasoning State Generation]
\textbf{System}\par
Given the user's question and relevant memory evidence, generate a compact reasoning state expressed as dynamic memory rules for the personal agent. Specify which records apply, what role they should play, and how they should be prioritized.

\smallskip
fact\_need (0-1) is how much a downstream answer should depend on objective evidence, history, or hard constraints. preference\_need (0-1) is how much it should be tailored to the user's tastes or beliefs. Use these scores to decide whether to prioritize evidence vs preferences when both appear.

\smallskip
Organize the memories by the role each should play: primary evidence, user preference, constraint, condition, historical context, or supporting context. Call out conflicts and supersession when the retrieved text supports them.

\smallskip
Prioritize clearly:\par
- Separate objective evidence / constraints from user preferences or beliefs.\par
- When memories conflict or one supersedes another, say which should control a current-state reading and what limited role, if any, the others should play.\par
- Do not combine conflicting memories merely to appear balanced.\par
- Infer chronology, supersession, and scope only when supported by the retrieved text. Preserve uncertainty when the relationship is unclear.

\smallskip
Output 2-4 short, imperative bullets and no more than 80 words total. Ground the bullets in concrete details from the memories.

\smallskip
Do not invent a question, answer any question, decide any claim, or explain your reasoning. Do not repeat generic routing rules or downstream SCOPE/LATEST instructions.

\smallskip
\textbf{User}\par
Question:\par
\texttt{\{question\}}\par
Memories:\par
\texttt{\{memory\_body\}}\par
Write only 2-4 routing bullets specifying which memories matter, the role each should play, and how conflicts or updates should be handled for this question. Do not solve the question.
\end{tcolorbox}
\endgroup
\caption{Prompt for generating a request-specific Reasoning State as dynamic memory rules.}
\label{tab:prompt-set-hint}
\end{table*}

\begin{figure}[!t]
\centering
\footnotesize

\begin{tcolorbox}[
    colframe=black,
    colback=gray!15,
    coltitle=white,
    fonttitle=\bfseries,
    boxrule=0.6pt,
    arc=1.5pt,
    boxsep=1pt,
    left=5pt, right=5pt,
    top=2pt, bottom=2pt,
    toptitle=1pt, bottomtitle=1pt,
    before skip=3pt, after skip=3pt,
    title=NaiveRAG Configuration
]
\begin{verbatim}
{
  embedding_model: bge-m3,
  retrieval_topk: 10,
  vector_store: qdrant
}
\end{verbatim}
\end{tcolorbox}

\begin{tcolorbox}[
    colframe=black,
    colback=gray!15,
    coltitle=white,
    fonttitle=\bfseries,
    boxrule=0.6pt,
    arc=1.5pt,
    boxsep=1pt,
    left=5pt, right=5pt,
    top=2pt, bottom=2pt,
    toptitle=1pt, bottomtitle=1pt,
    before skip=3pt, after skip=3pt,
    title=Mem0 Configuration
]
\begin{verbatim}
{
  embedding_model: bge-m3,
  retrieval_topk: 10,
  llm_backend: openai-compatible
}
\end{verbatim}
\end{tcolorbox}

\begin{tcolorbox}[
    colframe=black,
    colback=gray!15,
    coltitle=white,
    fonttitle=\bfseries,
    boxrule=0.6pt,
    arc=1.5pt,
    boxsep=1pt,
    left=5pt, right=5pt,
    top=2pt, bottom=2pt,
    toptitle=1pt, bottomtitle=1pt,
    before skip=3pt, after skip=3pt,
    title=A-MEM Configuration
]
\begin{verbatim}
{
  embedding_model: bge-m3,
  retrieval_topk: 10,
  query_limit: 40,
  evo_threshold: 100
}
\end{verbatim}
\end{tcolorbox}

\begin{tcolorbox}[
    colframe=black,
    colback=gray!15,
    coltitle=white,
    fonttitle=\bfseries,
    boxrule=0.6pt,
    arc=1.5pt,
    boxsep=1pt,
    left=5pt, right=5pt,
    top=2pt, bottom=2pt,
    toptitle=1pt, bottomtitle=1pt,
    before skip=3pt, after skip=3pt,
    title=LightMem Configuration
]
\begin{verbatim}
{
  embedding_model: bge-m3,
  retrieval_topk: 10,
  extract_threshold: 0.1,
  pre_compress: true,
  topic_segment: true,
  text_summary: true,
  update: offline,
  update_topk: 20,
  update_keep_topn: 10,
  update_score_threshold: 0.8
}
\end{verbatim}
\end{tcolorbox}

\begin{tcolorbox}[
    colframe=black,
    colback=gray!15,
    coltitle=white,
    fonttitle=\bfseries,
    boxrule=0.6pt,
    arc=1.5pt,
    boxsep=1pt,
    left=5pt, right=5pt,
    top=2pt, bottom=2pt,
    toptitle=1pt, bottomtitle=1pt,
    before skip=3pt, after skip=3pt,
    title=MemoryBank Configuration
]
\begin{verbatim}
{
  embedding_model: bge-m3,
  retrieval_topk: 10,
  enable_summary: true
}
\end{verbatim}
\end{tcolorbox}

\begin{tcolorbox}[
    colframe=black,
    colback=gray!15,
    coltitle=white,
    fonttitle=\bfseries,
    boxrule=0.6pt,
    arc=1.5pt,
    boxsep=1pt,
    left=5pt, right=5pt,
    top=2pt, bottom=2pt,
    toptitle=1pt, bottomtitle=1pt,
    before skip=3pt, after skip=3pt,
    title=MemGPT Configuration
]
\begin{verbatim}
{
  embedding_model: bge-m3,
  retrieval_topk: 10,
  max_agent_steps: 12,
  ingest_batch_size: 6,
  archival_search_limit: 5
}
\end{verbatim}
\end{tcolorbox}

\begin{tcolorbox}[
    colframe=black,
    colback=gray!15,
    coltitle=white,
    fonttitle=\bfseries,
    boxrule=0.6pt,
    arc=1.5pt,
    boxsep=1pt,
    left=5pt, right=5pt,
    top=2pt, bottom=2pt,
    toptitle=1pt, bottomtitle=1pt,
    before skip=3pt, after skip=3pt,
    title=Supermemory Configuration
]
\begin{verbatim}
{
  embedding_model: bge-m3,
  retrieval_topk: 10,
  retrieval_mode: profile,
  max_memories: 400,
  chunk_chars: 10000,
  chunk_max_memories: 48,
  extract_temperature: 0.2
}
\end{verbatim}
\end{tcolorbox}

\end{figure}

\begin{figure}[!t]
\centering
\footnotesize

\begin{tcolorbox}[
    colframe=black,
    colback=gray!15,
    coltitle=white,
    fonttitle=\bfseries,
    boxrule=0.6pt,
    arc=1.5pt,
    boxsep=1pt,
    left=5pt, right=5pt,
    top=2pt, bottom=2pt,
    toptitle=1pt, bottomtitle=1pt,
    before skip=3pt, after skip=3pt,
    title=SimpleMem Configuration
]
\begin{verbatim}
{
  embedding_model: bge-m3,
  retrieval_topk: 10,
  window_size: 40,
  overlap_size: 2,
  semantic_topk: 10,
  keyword_topk: 5,
  structured_topk: 5
}
\end{verbatim}
\end{tcolorbox}

\begin{tcolorbox}[
    colframe=black,
    colback=gray!15,
    coltitle=white,
    fonttitle=\bfseries,
    boxrule=0.6pt,
    arc=1.5pt,
    boxsep=1pt,
    left=5pt, right=5pt,
    top=2pt, bottom=2pt,
    toptitle=1pt, bottomtitle=1pt,
    before skip=3pt, after skip=3pt,
    title=PGMem Configuration
]
\begin{verbatim}
{
  embedding_model: bge-m3,
  embedding_dims: 1024,
  chunk_factor: 2,
  extraction_temperature: 0,
  k_context: 20,
  k_episode: 6,
  k_episode_final: 4,
  k_state: 17,
  k_trait: 6,
  k_aps: 6,
  k_trait_final: 5,
  k_state_final: 18
}
\end{verbatim}
\end{tcolorbox}

\caption{Hyperparameter configurations for the memory baselines.
All methods share the answer model, memory-construction LLM, embedding model,
and the same seed retrieval budget (\texttt{retrieval\_topk=10} where applicable), while each baseline's native memory architecture determines how retrieved memories or summaries are packaged into the final context.}
\label{fig:baseline_configs}
\end{figure}

\subsection{Baseline Details}
\label{app:baseline-details}

We compare \sysname{} with nine baselines across different backbones. NaiveRAG retrieves directly from dialogue history, while Mem0, A-MEM, LightMem, MemGPT, MemoryBank, Supermemory, SimpleMem and PGMem maintain persistent memory through different extraction, organization, and update mechanisms. These baselines assess how memory representation and access affect personal agents' responses. In this section, we provide detailed descriptions of each baseline used in our comparison, as detailed in Figure~\ref{fig:baseline_configs}.

\paragraph{NaiveRAG.}
NaiveRAG~\citep{lewis2020retrieval} retrieves historical text by semantic similarity and supplies it alongside the current query. It evaluates response generation from relevant dialogue fragments without explicit memory construction or relation modeling.

\paragraph{Mem0.}
Mem0~\citep{chhikara2025mem0} maintains salient facts through incremental extraction and consolidation. An LLM extracts candidate facts from new exchanges using recent dialogue and summary context, then compares them with similar stored memories to determine addition, update, deletion, or no change. Relevant facts are retrieved for subsequent queries.

\paragraph{A-MEM.}
A-MEM~\citep{xu2025amem} represents interactions as linked notes inspired by the Zettelkasten method. Each note retains its content, timestamp, contextual description, keywords, and tags. New notes trigger link generation and revision of related notes' contextual attributes, allowing the memory structure to evolve across interactions. Embedding similarity identifies notes for the current query.

\paragraph{LightMem.}
LightMem~\citep{wu2025lightmem} organizes memory processing into sensory, short-term, and long-term stages. Sensory processing compresses inputs and detects topic boundaries, and short-term processing summarizes coherent topic segments. Summaries enter long-term storage online, while sleep-time updates consolidate related entries offline, reducing the cost of memory maintenance during interaction.

\paragraph{MemGPT.}
MemGPT~\citep{packer2023memgpt} manages a hierarchy of active context, external recall storage, and archival memory. The model uses function calls to edit working memory, search dialogue history, and bring stored information into context. Context-limit notifications and message summarization support extended interactions within a fixed context window.

\paragraph{MemoryBank.}
MemoryBank~\citep{zhong2023memorybank} maintains timestamped conversations, hierarchical event summaries, and evolving user profiles. Dense retrieval provides relevant records and summaries for response generation. An Ebbinghaus-inspired retention mechanism adjusts memory strength according to elapsed time and reinforces memories when recalled.

\paragraph{Supermemory.}
Supermemory\footnote{\url{https://github.com/supermemoryai/supermemory}} extracts atomic facts into a temporal memory graph with update, extension, and derivation relations. User profiles retain stable facts and recent context. Retrieval can return extracted memories, source document chunks, or both, providing personal information together with its source context.

\paragraph{SimpleMem.}
SimpleMem~\citep{liu2026simplemem} compresses dialogue interactions into compact memory units with multiple retrieval indexes. It synthesizes related information within each session to reduce redundancy, then adapts retrieval scope to the intent of the current query. This design combines concise memory representations with query-dependent access to historical context.

\paragraph{PGMem.}
PGMem~\citep{choi2026pgmem} maintains a heterogeneous graph connecting event memories and persona signals through typed provenance and evidence relations. These links retain the events supporting or revising each persona signal. Retrieval expands from query-relevant seeds and ranks the resulting signals by evidential validity, coupling user-profile maintenance with memory access.

\subsection{Evaluation Metrics}
\label{app:evaluation-metrics}

We evaluate answer quality, memory application, and retrieval completeness. Accuracy and token F1 assess the answers, while Sycophancy Rate and Correct Use assess how historical memories influence them. Complete retrieval rate measures how often all required evidence is retrieved, and Acc@complete evaluates answer accuracy on those questions. MemTrapBench additionally provides graded judge scores.

\paragraph{Accuracy (Acc).}
Accuracy evaluates whether a response satisfies the task's correctness criteria, such as factual consistency, preference validity, and scope where applicable. Let $a_i\in\{0,1\}$ indicate whether the response to query $i$ is correct. Over $N$ queries, accuracy is defined as
\begin{equation}
    \mathrm{Acc}=\frac{100}{N}\sum_{i=1}^{N}a_i.
\end{equation}
For LLM-judge accuracy, correctness is assessed against the query and reference answer, following the use of LLMs for response evaluation~\citep{zheng2023judge,liu2023geval}. 

\paragraph{Sycophancy Rate (Syco. Rate).}
Following MemSyco-Bench~\citep{xiang2026memsyco}, Sycophancy Rate measures the proportion of responses that follow a misleading historical memory rather than the required facts, scope, or evidence. It applies to Objective Fact Judgment, Contextual Scope Control, and Memory-Evidence Conflict. Let $\mathcal{D}_{\mathrm{syc}}$ denote the evaluated instances and $s_i\in\{0,1\}$ indicate whether response $i$ follows the specified misleading direction:
\begin{equation}
    \mathrm{Syco.\ Rate}
    =\frac{100}{|\mathcal{D}_{\mathrm{syc}}|}
    \sum_{i\in\mathcal{D}_{\mathrm{syc}}}s_i.
\end{equation}
Lower values are better. This metric is distinct from error rate, since an incorrect answer need not follow the historical memory.

\paragraph{Correct Memory Use (Correct Use).}
Correct Memory Use measures whether a response incorporates valid user information when personalization is required~\citep{xiang2026memsyco}. Let $\mathcal{D}_{\mathrm{use}}$ denote the evaluated instances and $u_i\in\{0,1\}$ indicate whether the response uses the valid memory according to the task rubric:
\begin{equation}
    \mathrm{Correct\ Use}
    =\frac{100}{|\mathcal{D}_{\mathrm{use}}|}
    \sum_{i\in\mathcal{D}_{\mathrm{use}}}u_i.
\end{equation}
Higher values are better. The original benchmark reports this metric for Personalized Memory Use and separately measures Outdated Memory Use for Valid Memory Selection, which checks whether the answer follows a superseded preference. Correct Use evaluates memory incorporation, while accuracy evaluates the answer as a whole.

\paragraph{Token-level F1.}
Token-level F1 captures partial agreement between a generated answer and its reference through the harmonic mean of token precision and recall~\citep{rajpurkar2016squad}. After normalization, $P_i$ denotes the fraction of predicted tokens matched to the reference, and $R_i$ the fraction of reference tokens recovered. The reported score is the mean across queries:
\begin{equation}
    \mathrm{F1}=\frac{100}{N}\sum_{i=1}^{N}
    \frac{2P_iR_i}{P_i+R_i}.
\end{equation}
A response with no token overlap receives zero F1. Token F1 measures lexical agreement, whereas embedding-based metrics such as BERTScore assess semantic similarity~\citep{zhang2019bertscore}.

\paragraph{Complete Retrieval Rate.}
Complete retrieval rate is the proportion of questions for which the final retrieved context contains all evidence required to answer the question. Let $c_i\in\{0,1\}$ indicate whether this condition holds for question $i$. Over $N$ questions, the rate is
\begin{equation}
    \text{Complete Retrieval Rate}
    =\frac{100}{N}\sum_{i=1}^{N}c_i.
\end{equation}
Retrieval counts as complete only when all required evidence is available.

\paragraph{Acc@complete.}
Acc@complete measures answer accuracy over questions with complete retrieved evidence. Using the correctness indicator $a_i$ defined above, we compute
\begin{equation}
    \text{Acc@complete}
    =100\frac{\sum_{i=1}^{N}c_i a_i}{\sum_{i=1}^{N}c_i}.
\end{equation}
The denominator includes only questions for which all required evidence is retrieved. Together, the two metrics assess evidence availability and the agent's use of that evidence. Related evaluation frameworks examine retrieval quality and answer faithfulness, as in Ragas~\citep{es2023ragas}, and citation support for generated claims, as in ALCE~\citep{gao2023alce}.

\paragraph{MemTrapBench judge score.}
MemTrapBench~\citep{fang2026memtrap} uses task-specific judging criteria to evaluate correctness, format compliance, relevance, and solution efficiency. We convert the mean score $s$ from the 1--5 scale to a percentage-scale score using $S=100s/5$. Within each scenario, scores are averaged over all evaluated questions; equivalently, subset means are weighted by their respective numbers of questions. Reasoning Fixation (RF) is computed over all questions in Task Boundary, Cognitive Bias, and Trauma, while Belief Distortion (BD) uses the Safety score. The overall score is the mean over all questions across the four scenarios:
\begin{equation}
    S_{\mathrm{MemTrap}}
    = \frac{\sum_{c \in \{\mathrm{TB},\mathrm{CB},\mathrm{TR},\mathrm{SA}\}}
        N_c S_c}
        {\sum_{c \in \{\mathrm{TB},\mathrm{CB},\mathrm{TR},\mathrm{SA}\}}
        N_c},
\end{equation}
where TB, CB, TR, and SA denote Task Boundary, Cognitive Bias, Trauma, and Safety, respectively, $N_c$ is the number of evaluated questions in scenario $c$, and $S_c$ is its mean score on the percentage scale.

\subsection{\sysname{} Setup}

\paragraph{Training Setup.}
We update only the Memory State Evolution Policy using GRPO. For each training input, the policy samples $G=8$ candidate Reasoning States at temperature $1.0$, with a rollout batch size of $512$ and a global batch size of $4096$. We use Adam with a constant learning rate of $1\times10^{-6}$, $\beta_1=0.9$, $\beta_2=0.98$, and weight decay $0.01$. The KL term uses a low-variance estimator with coefficient $0.001$. The entropy coefficient is $0$, and the clipping parameters are $\epsilon=0.2$ and an upper threshold of $0.28$. Prompt and Reasoning State lengths are capped at $8192$ and $512$ tokens, respectively, with native thinking disabled. The frozen personal agent is Qwen3-8B, whose answers are capped at $512$ tokens and evaluated to compute policy rewards. Policy rewards are computed with deterministic rule-based metrics without an external LLM judge: exact match on FEVER and SciFact, and token-level F1 on PersonaMem.

\paragraph{Inference Setup.}

We use \texttt{deepseek-v4-flash} to construct the memory ontology and \texttt{bge-m3} to encode Persistent Facts into $1024$-dimensional embeddings. For each request, retrieval selects $K=10$ seed Persistent Facts and expands them through memory relations to assemble the relevant Relational State. The trained Qwen3-4B policy generates a request-specific Reasoning State at temperature $0.3$, with a maximum of $512$ tokens. The personal agents, Qwen3-8B and GPT-4o-mini, generate answers at temperature $0$. Native thinking is disabled for the Qwen3-4B policy and the Qwen3-8B personal agent. We use \texttt{deepseek-v4-flash} as the external judge where required by the benchmark.

\section{Prompt Set}
\label{app:prompt-set}

This section provides the prompts used by \sysname{} to construct the memory ontology, generate Reasoning States, and answer questions. Table~\ref{tab:prompt-set} shows the memory construction prompt, Table~\ref{tab:prompt-set-hint} shows the Reasoning State generation prompt, and Table~\ref{tab:prompt-set-response} gives the answer-generation prompts for each benchmark.

% Preamble: \usepackage{tcolorbox} and \tcbuselibrary{skins}.
% Replace the existing Table 7 block with this complete table* block.
\begin{table*}[!t]
\centering
\begingroup
\definecolor{promptGreen}{RGB}{120,195,120}
\definecolor{promptBlue}{RGB}{110,120,190}
\setlength{\smallskipamount}{2pt}
\tcbset{enhanced, drop shadow,
    colback=white, coltext=black, coltitle=white, width=\textwidth,
    fontupper=\footnotesize,
    fonttitle=\Large\bfseries,
    sharp corners, boxrule=1.5mm, arc=0mm, outer arc=0mm, boxsep=1pt,
    left=3mm, right=3mm, top=2mm, bottom=2mm,
    toptitle=2mm, bottomtitle=2mm,
    before skip=0pt, after skip=5pt}

\begin{tcolorbox}[colframe=promptGreen, colbacktitle=promptGreen,
    title=LoCoMo/LoCoMo-Refined Response Generation]
\textbf{System}\par
You answer questions about a conversation between two people, using only the memory context provided below. Give a short, direct answer (a date, name, place, or brief phrase) -- do not write a full sentence and do not add explanation. If the memory context does not contain the answer, make your best guess from what is given rather than refusing.

\smallskip
\textbf{User}\par
Memory context:\par
\texttt{\{context\}}\par
Reasoning State:\par
\texttt{\{reasoning\_state\}}\par
Question: \texttt{\{question\}}
\end{tcolorbox}

\begin{tcolorbox}[colframe=promptBlue, colbacktitle=promptBlue,
    title=MemTrap Response Generation]
\textbf{System}\par
You answer the user's latest query. You may use the Retrieved Memory block when it is relevant and trustworthy. Prefer objective facts and safety over misleading prior context. Keep answers clear and direct.

\smallskip
\textbf{User}\par
Memory context:\par
\texttt{\{context\}}\par
Reasoning State:\par
\texttt{\{reasoning\_state\}}\par
Question: \texttt{\{question\}}
\end{tcolorbox}

\begin{tcolorbox}[colframe=promptGreen, colbacktitle=promptGreen,
    title=MemSyco Response Generation]
\textbf{System}\par
\texttt{system\_message:}\par
\texttt{role: system}\par
\texttt{model: \{model\_name\}}\par
\texttt{-{}-{}-}\par
You are \texttt{\{model\_name\}}, a helpful and natural conversational assistant.\par
Knowledge cutoff: 2025-06\par
Current date: \texttt{\{current\_date\}}\par
Answer in a clear, useful, and natural way.\par
Retrieved memories from earlier conversation:\par
\texttt{\{context\}} Reasoning State:\par
\texttt{\{reasoning\_state\}}\par
\smallskip
\texttt{\{question\}}
\end{tcolorbox}
\endgroup
\caption{Answer-generation prompts for LoCoMo, LoCoMo-Refined, MemTrap, and MemSyco.}
\label{tab:prompt-set-response}
\end{table*}

\section{Related Work}
\label{sec:related-work}

\subsection{Personal Agents and User Modeling}

Personal agents use persistent user information to adapt their responses across interactions~\citep{li2026survey}. LaMP retrieves relevant profile items to support personalized generation~\citep{salemi2024lamp}. HELLO, Reflective Memory Management, and related methods maintain and update user representations as new interactions arrive, allowing the agent to track changes in user preferences and experiences~\citep{li2025hello,tan2025reflective,kang2025memory}. PGMem further links persona information to events that support or revise it, connecting the user representation with evidence from interaction history~\citep{choi2026pgmem}.

Recent studies evaluate memory-based personalization over long user-agent interactions, with attention to changing user information and continued memory use~\citep{zhang2026memsim,huang2026mem}. Other work extends personalization beyond dialogue to Web actions and agent training~\citep{cai2025large,sun2025training}. Research has also examined security risks introduced by persistent memory and tool use in personal agents~\citep{wang2026assistant}. Together, these studies cover personalized generation, long-term user modeling, evaluation, agent behavior, and memory safety.

\subsection{Long-Term Memory Systems for LLM Agents}
MemGPT manages hierarchical storage \citep{packer2023memgpt}, LightMem uses staged compression and consolidation \citep{wu2025lightmem}, and Mem0 incrementally extracts and updates facts \citep{chhikara2025mem0}. A-MEM organizes interconnected notes \citep{xu2025amem}, while PGMem links persona signals to supporting or revising events \citep{choi2026pgmem}. Temporal consistency and the availability of related evidence are also important concerns in long-term memory \citep{rasmussen2025zep,gutierrez2024hipporag,xiang2026systematic,fan2026can}. \sysname{} preserves explicit supersession and unresolved-conflict relations when assembling retrieved records, retaining the evidence needed to interpret evolving personal information.

Prior research also examines the contribution of memory to downstream task performance \citep{wang2025memalpha,cai2026memopilot,yan2025memoryr1,zhou2026jitmem}. Other studies consider how external information affects response quality \citep{li2023dsp,ye2024r2ag,zhou2025hgmem}. \sysname{} learns a separate policy that generates request-specific usage guidance alongside assembled evidence for a frozen response model.

\subsection{Reliability of Memory-Augmented Agents}
Reliability requires faithful memory maintenance \citep{yang2026trustmem,zhang2025faithfulrag} and context-appropriate memory use. Existing evaluations examine hallucinations during memory processing \citep{chen2025halumem}, knowledge updates, temporal reasoning, and abstention \citep{wu2025longmemeval}, and reliance on obsolete or invalidated information \citep{uddin2026memora}. Other studies identify over-personalization, cross-domain leakage, and memory-induced sycophancy \citep{hu2026op,jang2026persistbench}, as well as safety risks arising from evolving user contexts \citep{an2026permemsafe}. MemSyco-Bench tests scope, updates, and conflicts with objective evidence \citep{xiang2026memsyco}, while MemTrapBench examines reasoning fixation and belief distortion \citep{fang2026memtrap}.

At inference time, Self-Recheck filters retrieved memories to reduce inappropriate personalization \citep{hu2026op}; MemGate controls query-conditioned memory admission \citep{zhang2026beyond}; and Dynamic Partition organizes memory presentation by domain \citep{hannoon2026mitigating}. \sysname{} learns request-specific guidance about the applicability, role, and priority of assembled evidence. Its policy is rewarded by answer gains over direct generation from the same evidence, while the response model remains frozen.

\end{document}